\documentclass[twocolumn, times]{aastex63}

\usepackage[english]{babel}
\usepackage{newtxtext}
\usepackage[T1]{fontenc}
\usepackage{ae, aecompl}
\usepackage{etoolbox}
\usepackage{amsmath, amssymb}
\usepackage{graphicx, mathptm, mathrsfs} 
\usepackage{times} 
\usepackage{natbib}
\usepackage{bigints}
\usepackage{color, xcolor, colortbl}
\usepackage{soul}
\usepackage{booktabs}
\usepackage{epstopdf}
\usepackage[title]{appendix}

\usepackage{scalerel}
\usepackage{tikz}
\usetikzlibrary{svg.path}

\def \apjl{Astrophys. J. Lett.}
\def \apjs{Astrophys. J. Suppl.}

\def \aap{Astron. Astrophys.}

\newcommand{\Mpch}{$h^{-1}\,\mbox{Mpc}$\,}

\defcitealias{wen2012}{WHL12}
\defcitealias{Planck2018}{Planck18}
  
\submitjournal{ApJ}

\shorttitle{Constraints on the linear growth rate from cluster
  clustering}

\shortauthors{F. Marulli al.}

\graphicspath{{./}{Figures/}}

\begin{document}

\title{C$^3$ - Cluster Clustering Cosmology \\I. New constraints on
  the cosmic growth rate at $z\sim0.3$ from redshift-space clustering
  anisotropies}

\correspondingauthor{Federico Marulli}
\email{federico.marulli3@unibo.it}

\author[0000-0002-8850-0303]{Federico Marulli}
\affiliation{Dipartimento di Fisica e Astronomia ``Augusto Righi'' -
  Alma Mater Studiorum Universit\`{a} di Bologna, via Piero Gobetti
  93/2, I-40129 Bologna, Italy}
\affiliation{INAF - Osservatorio di Astrofisica e Scienza dello Spazio
  di Bologna, via Piero Gobetti 93/3, I-40129 Bologna, Italy}
\affiliation{INFN - Sezione di Bologna, viale Berti Pichat 6/2,
  I-40127 Bologna, Italy}

\author[0000-0003-2387-1194]{Alfonso Veropalumbo}
\affiliation{Dipartimento di Fisica, Universit\`{a} degli Studi Roma
  Tre, via della Vasca Navale 84, I-00146 Rome, Italy}
\affiliation{INFN - Sezione di Roma Tre, via della Vasca Navale 84,
  I-00146 Rome, Italy}

\author[0000-0001-6667-5471]{Jorge Enrique Garc\'ia-Farieta}
\affiliation{Center for Theoretical Physics, Polish Academy of
  Sciences, Al. Lotników 32/46, 02-668 Warsaw, Poland}
\affiliation{Departamento de F\'isica, Universidad Nacional de
  Colombia - Sede Bogot\'a, Av. Cra 30 No 45-03, Bogot\'a, Colombia}
\affiliation{Dipartimento di Fisica e Astronomia ``Augusto Righi'' -
  Alma Mater Studiorum Universit\`{a} di Bologna, via Piero Gobetti
  93/2, I-40129 Bologna, Italy}

\author[0000-0002-7616-7136]{Michele Moresco}
\affiliation{Dipartimento di Fisica e Astronomia ``Augusto Righi'' -
  Alma Mater Studiorum Universit\`{a} di Bologna, via Piero Gobetti
  93/2, I-40129 Bologna, Italy}
\affiliation{INAF - Osservatorio di Astrofisica e Scienza dello Spazio
  di Bologna, via Piero Gobetti 93/3, I-40129 Bologna, Italy}

\author[0000-0002-3473-6716]{Lauro Moscardini}
\affiliation{Dipartimento di Fisica e Astronomia ``Augusto Righi'' -
  Alma Mater Studiorum Universit\`{a} di Bologna, via Piero Gobetti
  93/2, I-40129 Bologna, Italy}
\affiliation{INAF - Osservatorio di Astrofisica e Scienza dello Spazio
  di Bologna, via Piero Gobetti 93/3, I-40129 Bologna, Italy}
\affiliation{INFN - Sezione di Bologna, viale Berti Pichat 6/2,
  I-40127 Bologna, Italy}

\author[0000-0002-4409-5633]{Andrea Cimatti}
\affiliation{Dipartimento di Fisica e Astronomia ``Augusto Righi'' -
  Alma Mater Studiorum Universit\`{a} di Bologna, via Piero Gobetti
  93/2, I-40129 Bologna, Italy}
\affiliation{INAF - Osservatorio Astrofisico di Arcetri, Largo
  Enrico Fermi 5, I-50125, Florence, Italy}

\begin{abstract}
  
Redshift-space distortions in the clustering of galaxy clusters
provide a novel probe to test the theory of gravity on cosmological
scales. The aim of this work is to derive new constraints on the
linear growth rate of cosmic structures from the redshift-space
two-point correlation function of galaxy clusters. We construct a
large spectroscopic catalog of optically-selected clusters from the
Sloan Digital Sky Survey. The selected sample consists of $43,743$
clusters in the redshift range $0.1<z<0.42$, with masses estimated
from weak-lensing calibrated scaling relations. We measure the
transverse and radial wedges of the two-point correlation function of
the selected clusters. Modeling the redshift-space clustering
anisotropies, we provide the first constraints on the linear growth
rate from cluster clustering. The cluster masses are used to set a
prior on the linear bias of the sample. This represents the main
advantage in using galaxy clusters as cosmic probes, instead of
galaxies. Assuming a standard cosmological model consistent with the
latest cosmic microwave background constraints, we do not find any
evidence of deviations from general relativity. Specifically, we get
the value of the growth rate times the matter power spectrum
normalization parameter $f\sigma_{8}=0.44\pm0.05$, at an effective
redshift of $z=0.275$.

\end{abstract}

\keywords{Observational cosmology -- Cosmological parameters -- Redshift surveys -- Galaxy clusters}


\section{Introduction}

The spatial distribution of matter in the universe depends on both the
expansion rate of space and the peculiar velocities at small scales
caused by local gravitational interactions. Second-order and
third-order summary statistics of the matter density field, i.e. the
two-point (2PCF) and three-point auto-correlation functions, provide
key information on the main cosmological model parameters, and can be
effectively assessed through the corresponding statistics of properly
selected samples of biased cosmic tracers, such as galaxies. In
particular, apparent anisotropies in the 2PCF, induced by neglecting
the peculiar velocities along the line of sight when computing
comoving distances, can be effectively exploited to test the gravity
theory on the largest cosmological scales. Redshift-space distortions
\citep[RSD,][]{kaiser1987, hamilton1998} in clustering statistics
provide an indirect measurement of the properties of the matter
peculiar velocity field, which can be parameterized by the linear
growth rate of cosmic structures, $f\equiv d\log G/d\log a$, where $G$
is the growth factor and $a$ is the scale factor \citep{peacock2001,
  hawkins2003, guzzo2008, zhang2008}. Combining measurements of the
cosmic growth rate and of the Hubble expansion rate it is possible to
discriminate among alternative dark energy models \citep{linder2017,
  moresco2017}.

Large and dense samples of extragalactic sources are required to
accurately measure the 2PCF in a wide enough range of comoving
coordinates and redshifts, at sufficiently high signal-to-noise
ratio. Different tracers are generally considered to maximize the
redshift range covered \citep[e.g.][]{eBOSS2021}. Measurements from
the auto- and cross-correlation functions of galaxies
\citep[e.g.][]{percival2004, samushia2012, samushia2014, tojeiro2012,
  reid2012, chuang2013, chuang2013b, chuang2016, beutler2014,
  okumura2016, delatorre2017, adams2017, pezzotta2017, mohammad2018,
  icaza-lizaola2020, wang2020}, quasars \citep[e.g.][]{neveux2020,
  hou2021}, cosmic voids \citep[e.g.][]{hamaus2016, hamaus2020,
  hawken2017, hawken2020, nadathur2019, nadathur2020, aubert2020} and
other probes \citep[e.g.][]{davis2011, turnbull2012, hudson2012,
  feix2015} allowed testing the gravity theory on a wide redshift
range, up to $z\sim1.5$ (see the discussion of our results in \S
\ref{sec:Results}).

The goal of this work is to provide new constraints on the linear
growth rate of cosmic structures from the redshift-space 2PCF of a
large spectroscopic sample of galaxy clusters extracted from the Sloan
Digital Sky Survey (SDSS), at an effective redshift $z\sim0.3$. In
\citet{moresco2021} we analyze the three-point correlation function of
the same catalog up to the baryon acoustic oscillations (BAO) scales
and provide constraints on the nonlinear bias of the sample, while in
\citet{veropalumbo2021} we perform a joint RSD+BAO analysis of the
two-point and three-point correlation functions.

Galaxy clusters are the biggest structures that are virialized in the
present universe. Large-scale cluster statistics provide one of the
primary probes to constrain the universe's geometry and growth rate,
especially because the masses of dark matter haloes hosting clusters
can be accurately assessed via different techniques, exploiting the
cluster multiwavelength signal. In particular, the redshift evolution
of cluster number counts provides strong cosmological constraints on
the total matter energy density parameter, $\Omega_{\rm M}$, and on
the amplitude of the matter power spectrum, $\sigma_8$ \citep[see
  e.g.][and references therein]{vikhlinin2009, pacaud2018,
  costanzi2019, lesci2020}. The clustering of galaxy clusters is a
harder statistics to measure, as it requires dense samples of sources
in a wide comoving separation range. Nevertheless, the cluster 2PCF
has already been deeply exploited in cosmological studies, also in
combination with other cluster probes such as number counts and
gravitational lensing \citep[see e.g.][]{moscardini2000b, miller2001,
  schuecker2001, schuecker2003, majumdar2004, estrada2009, hutsi2010,
  balaguera2011, hong2012, hong2016, mana2013, veropalumbo2014,
  veropalumbo2016, sereno2015, emami2017, marulli2018, nanni2020}.

In fact, cluster clustering offers several key advantages relative to
galaxy clustering \citep[e.g.][]{angulo2005, marulli2017}. Galaxy
clusters are highly biased tracers; thus, at a given scale, their 2PCF
clustering signal is high compared to galaxies, and increases with
cluster masses \citep{sheth_mo_tormen2001}. Furthermore, pure enough
galaxy cluster samples, with a negligible fraction of satellite
galaxies erroneously identified as Brightest Cluster Galaxies (BCGs),
are relatively less affected by nonlinear dynamics at small scales --
the so-called Fingers-of-God distortions, which reduces the impact of
possible systematics from RSD model assumptions \citep{valageas2012,
  marulli2017}. Large spectroscopic cluster catalogs have also been
proven to be optimal probes for BAO cosmological analyses, due to low
damping in the BAO shape as compared to galaxy clustering
\citep{hong2012, hong2016, veropalumbo2014, veropalumbo2016}. On the
other hand, high-mass cluster-scale haloes are known to exhibit
nonlinear bias \citep[see e.g.][]{desjacques2018} which should be
properly modeled in the likelihood function, as discussed in \S
\ref{subsec:rsd}. Finally, as mentioned before, another key benefit of
using clusters as cosmological probes is the possibility of assessing
cluster masses, which can be used in cluster clustering analyses to
estimate the effective bias of the sample when a cosmological model is
assumed. Furthermore, the cosmological dependence of cluster mass
estimates might be exploited to further strengthen the cosmological
constraints.

Throughout this paper we assume a fiducial $\Lambda$-cold dark matter
($\Lambda$CDM) cosmological model consistent with
\citet[][\citetalias{Planck2018}]{Planck2018} parameters, i.e.
$\Omega_M = 0.3153$, $\Omega_\Lambda = 0.6846$, $\Omega_b = 0.0486$,
$\sigma_8 = 0.8111$, $n_s = 0.9649$. The dependence of observed
coordinates on the Hubble parameter is expressed as a function of
$h\equiv H_0/100\, {\rm km\, s^{-1} Mpc^{-1}}$.

The analyses presented in this work have been performed with the
\texttt{CosmoBolognaLib}\footnote{In this work we used the
  \texttt{CosmoBolognaLib V5.5}. The software is released at
  \href{https://gitlab.com/federicomarulli/CosmoBolognaLib}{gitlab.com/federicomarulli/CosmoBolognaLib},
  together with documentation and example codes.}
\citep{marulli2016}, a set of {\em free software} numerical libraries
that we used here to handle the catalog of galaxy clusters, measure
their clustering statistics, and perform Bayesian statistical analyses
aimed at extracting constraints from redshift-space clustering
anisotropies.

The paper is organized as follows. In \S \ref{sec:Data} we present the
spectroscopic cluster sample used in this work, describing the
selection criteria and cluster main properties. In \S
\ref{sec:ClusteringMeasurements} and \S \ref{sec:Modeling} we explain
the adopted methods and assumptions to measure and model the
redshift-space clustering wedges, respectively. The results of the
analysis are presented and discussed in \S \ref{sec:Results}, while \S
\ref{sec:Conclusions} summarizes the main findings of this work.

\begin{figure*}
  \centering
  \includegraphics[width=0.8\textwidth]{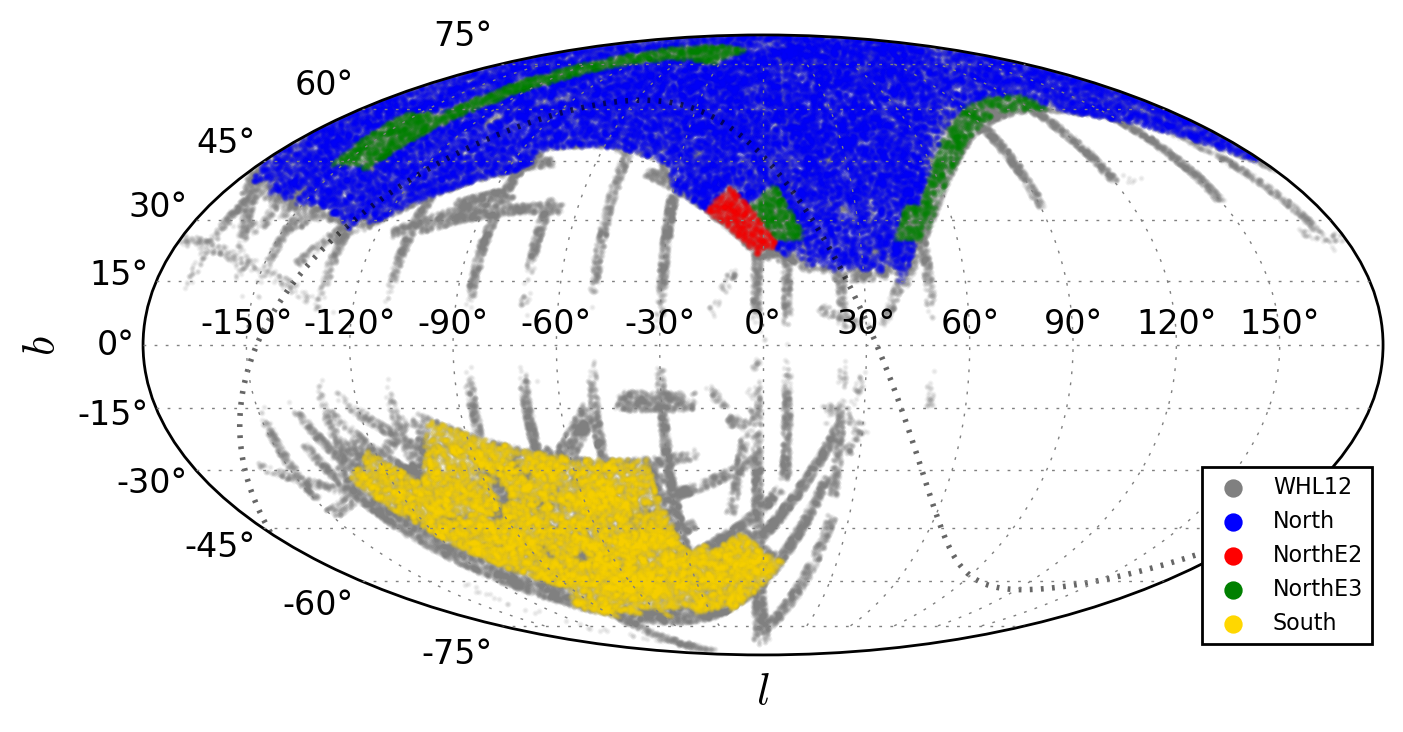}
  \caption{The angular distribution of the spectroscopic cluster
    sample analyzed in this work, in the north (blue points), north E2
    (red points), north E3 (green points) and south (yellow points)
    fields, compared to the \citetalias{wen2012} photometric sample
    (gray points, in background). The map is shown in the Galactic
    Coordinate System. The dotted black line indicates the Celestial
    Equator.}
  \label{fig:map}
\end{figure*}


\section{The data}
\label{sec:Data}


\subsection{The photometric sample}
\label{subsec:photometric}

The catalog analyzed in this work consists of optically selected
clusters of galaxies that have been identified by
\citet[][\citetalias{wen2012}]{wen2012}\footnote{The latest version of
  the \citetalias{wen2012} catalog is publicly available at
  \href{http://zmtt.bao.ac.cn/galaxy\_clusters/}{http://zmtt.bao.ac.cn/galaxy\_clusters}.}
from the Sloan Digital Sky Survey III \citep[][SDSS-III, Data Release
  (DR) 8]{aihara2011}.

The \citetalias{wen2012} catalog lists $132,684$ galaxy clusters on a
sky area of $\sim15,000$ square degrees, spanning the redshift range
$0.05 < z < 0.8$.  The cluster identification is based on a
friends-of-friends procedure \citep{huchra1982}. This approach has
been already exploited to find groups and clusters using
volume-limited spectroscopic samples of galaxies \citep[see
  e.g.][]{berlind2006,tempel2014} at low redshifts ($z<0.2$). The
\citetalias{wen2012} cluster sample extends the technique on
photometric redshift samples of galaxies, allowing the detection of
galaxy overdensities around the BCGs at higher redshifts.

A candidate cluster is included in the catalog if $N_{200}\ge 8$ and
$R_{L_*}\ge 12$, where $N_{200}$ is the number of member candidates
within $\tilde{r}_{200}$, and $R_{L_*}$ is the optical richness
defined as $R_{L_*}= \tilde{L}_{200} / L_*$, where $\tilde{L}_{200}$
is the total $r$-band luminosity within an empirically determined
radius $\tilde{r}_{200}$ and $L_*$ is the evolved characteristic
galaxy luminosity \citep{blanton2003}.  The subscript $200$ denotes
quantities measured in a sphere whose mean density is $200$ times the
critical density at the cluster redshift. The cluster photometric
redshifts reported in the catalog are the median value of the
photometric redshifts of the galaxy members. These selections have
been applied to avoid contaminations by bright field galaxies with
overestimated photometric redshifts.

The cluster masses are estimated from the weak-lensing cluster mass
scaling relation calibrated in \citetalias{wen2012}, with data from
\citet{wen2010}:
\begin{equation}
  \log\left(\frac{M_{200}}{10^{14}M_\odot}\right) = (-1.49 \pm 0.05) +
  (1.17 \pm 0.03) \log\left(R_{L_*}\right) \, .
  \label{eq:scaling}
\end{equation}
To verify the robustness of our analysis, we consider also the scaling
relation provided independently by \citet{covone2014}, finding
consistent results.

According to \citetalias{wen2012}, the catalog completeness, which is
the fraction of the selected galaxy clusters over the full sample, is
close to $1$ in the redshift range $0.1<z<0.42$, for
$M_{200}\gtrsim2\times10^{14} M_{\odot}$, while the detection rate
decreases down to $\sim75 \%$, including all clusters down to the
minimum mass of the sample, $M_{200} = 6\times10^{13} M_{\odot}$.
\citetalias{wen2012} also quantified the false cluster detections to
be at the level of $6\%$ for $R_{L_*} = 12$, decreasing to $<1\%$ for
cluster of richness $R_{L_*} \geq 23$. Possible effects on the
properties of the \citetalias{wen2012} cluster sample caused by
incompletenesses of SDSS-III's Baryon Oscillation Spectroscopic Survey
(BOSS) galaxies at high stellar masses \citep{leauthaud2016,
  saito2016} are neglected in this work. As we will detail in \S
\ref{sec:Modeling}, our statistical analysis does not depend on the
cluster catalog completeness.


\subsection{The spectroscopic sample}
\label{subsec:spectroscopic}

An accurate and precise estimate of the redshift is crucial when
reconstructing the statistical properties of the large-scale
distribution of matter. Large redshift uncertainties, as in
photometric redshift surveys, lead to severe distortion effects that
reflect in the 2PCF measurement, complicating its analysis and
cosmological interpretation \citep[see e.g.][]{marulli2012b,
  sereno2015, garcia-farieta2020}. In order to construct a
spectroscopic cluster sample, we take advantage of the spectroscopic
data from the SDSS, focusing on the final spectroscopic DR12 from BOSS
\citep{dawson2013, alam2015, alam2017}, which is part of the SDSS-III
program. This survey measured the spectra for millions of galaxies.
We assign spectroscopic redshifts to \citetalias{wen2012} clusters by
crossmatching with the spectroscopic galaxy sample\footnote{The match
  has been done using the {\small OBJID} entry.}. The total cluster
catalog with spectroscopic information consists of $72,563$ objects,
spanning the redshift range $0<z<1$. Following \citetalias{wen2012},
we cut the sample in the redshift range $0.1<z<0.42$, to minimize
incompleteness uncertainties. The number of remaining clusters is
$43,743$, with a median redshift of $z\sim0.3$, covering an area of
about $10,800\,\mbox{deg}^2$.

We make no distinction between galaxy clusters and BCGs in this
analysis, since the coordinates of galaxy clusters are estimated as
the coordinates of their BCGs, without further refinements.  Clusters
with no spectroscopic information for their BCGs are discarded, even
if some of their member galaxies have a measured spectroscopic
redshift. The rationale of this choice is to reduce contamination from
nonlinear dynamics in hosting virialized dark matter haloes, thus
minimizing the impact of theoretical uncertainties in the RSD
modeling at small scales.

Figure \ref{fig:map} shows the angular distribution of the
spectroscopic galaxy cluster catalog analyzed in this work, compared
to the \citetalias{wen2012} photometric sample. The three North fields
and the South one are shown with different colours. The early (E)
North fields (E2, E3) have been included in DR12. They are
characterized by a lower galaxy density and a different redshift
distribution, relative to the North and South fields
\citep{beutler2017}, and will be treated differently when constructing
the random catalog (see \S \ref{subsec:RandomCatalog}).

\begin{figure*}
  \includegraphics[width=0.49\textwidth]{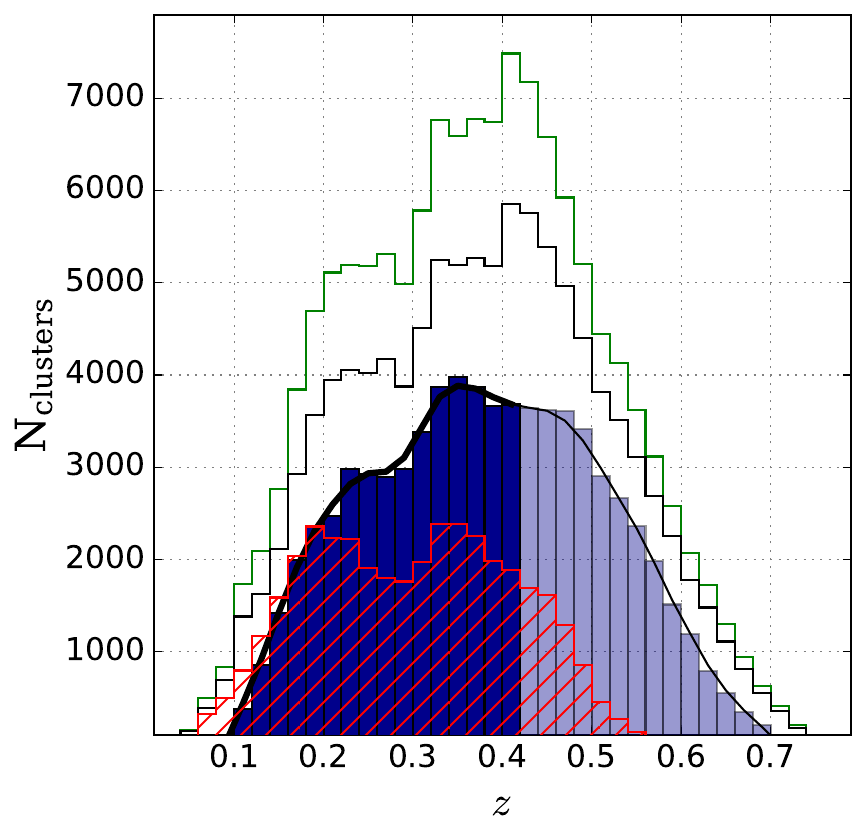}
  \includegraphics[width=0.49\textwidth]{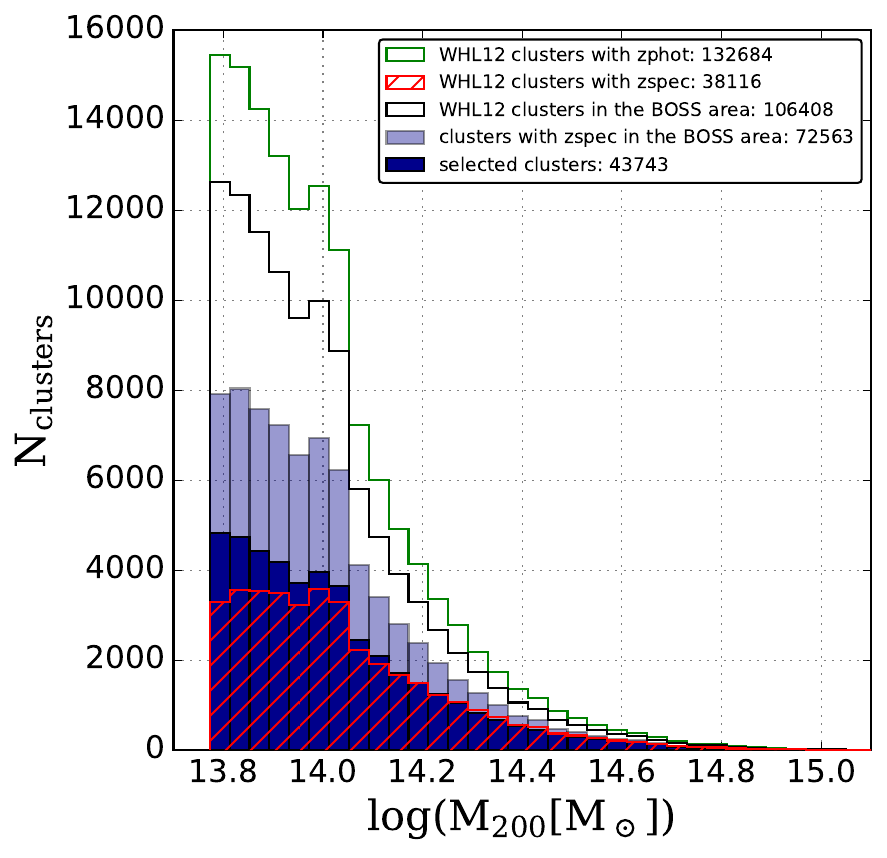}
  \caption{The redshift ({\em left panel}) and mass ({\em right
      panel}) distribution of the spectroscopic cluster sample
    analyzed in this work (blue solid histogram in the selected
    redshift range $0.1<z<0.42$, cyan histogram at higher redshifts),
    compared to the redshift distributions of the \citetalias{wen2012}
    photometric and spectroscopic cluster samples (green and red
    histogram, respectively) and of the \citetalias{wen2012} sample
    restricted to the BOSS area (black histogram). The number of
    selected clusters in each catalog is reported in the label. The
    black solid line in the left panel shows the smoothed redshift
    distribution of the random catalog.}
  \label{fig:distr}
\end{figure*}

Figure \ref{fig:distr} compares the redshift and mass distributions of
the spectroscopic cluster sample analyzed in this work to the ones of
the \citetalias{wen2012} photometric and spectroscopic cluster
catalogs, and of the \citetalias{wen2012} sample restricted to the
BOSS area. The shape of the mass distribution of the selected cluster
sample is overall consistent with theoretical $\Lambda$CDM predictions
by \citet{tinker2008}. However, we do not attempt to exploit the
cluster mass distribution in this work, to avoid systematics due to
possible inaccurate knowledge of the sample selection function. The
estimated masses are used instead to set a prior on the linear bias of
the selected cluster sample.

\section{Clustering measurements}
\label{sec:ClusteringMeasurements}

In this Section, we present the methodologies considered in this work
to measure the redshift-space wedges of the selected galaxy cluster
sample, that will be used to derive constraints on the linear growth
rate from RSD.


\subsection{Random catalog}
\label{subsec:RandomCatalog}

To estimate the three-dimensional 2PCF of a sample of extragalactic
sources, a geometric selection function is needed. The clustering
estimator adopted in this work (see \S \ref{subsec:wedges}) requires
this function to be provided as a catalog of objects randomly
distributed in the same area of the real catalog, and with the same
selection along the line of sight. As we will explain in \S
\ref{sec:Modeling}, the likelihood we will use to extract cosmological
constraints from the analyzed clustering dataset is independent of the
catalog completeness, which is the fraction of selected galaxy
clusters over the full sample \citep[see][for a similar
  analysis]{marulli2018}. Thus, our final results depend only on the
geometric selection that enters the clustering estimator.

We construct the random catalog following the same methodology used in
galaxy clustering analyses, in particular, the one used to measure the
2PCF of BOSS galaxies. In the assumption that the angular and redshift
distributions of the selected galaxy clusters are independent, we
assign angular coordinates, i.e. R.A., decl., and redshifts in two
separate steps. The R.A.-decl. coordinates are extracted with
\texttt{MANGLE} \citep{swanson2008_MANGLE}, using publicly available
survey
footprints\footnote{\href{https://data.sdss.org/sas/dr12/boss/lss/}{data.sdss.org/sas/dr12/boss/lss}}. The
redshifts are then sampled from the true redshift distribution of the
catalog, smoothed with a Gaussian kernel of $\sigma_z=0.02$ in order
not to introduce spurious clustering along the line of sight. As we
verified, the impact of this assumption is negligible. Due to the
different density and redshift distributions in the north, north E2,
north E3 and south fields, the extraction of random coordinates and
redshifts is performed in each field separately. The final random
catalog, which is obtained by adding the random catalogs in the four
fields, is constructed to be $50$ times larger than the selected
cluster catalog in order to minimize the impact of the shot noise. The
redshift distribution of the random objects normalized to the number
of the spectroscopic cluster sample is shown in Fig. \ref{fig:distr}.


\subsubsection{Weights}
\label{subsec:weights}
We take into account the systematic uncertainties in the angular
selection function due to the position-dependent completeness of the
cluster sample. To do this, we cross correlate the angular cluster
counts with the maps of observational systematics provided by
\citet{leistedt2014}\footnote{\href{https://www.earlyuniverse.org/release-of-the-sdss-systematics-templates/}{www.earlyuniverse.org/release-of-the-sdss-systematics-templates}}. We
find that the cluster counts are anticorrelated with stellar
densities, and less strongly, with the $r$-band extinction. We correct
for this effect by weighting the objects in the random sample
accordingly.

Moreover, we weight the clusters in the catalog to account for the
spectroscopic target selection, following the same weighting scheme
adopted by \citet{reid2016} and \citet{ross2017}:
\begin{equation}
  w = w_{\rm see} w_{\rm star} (w_{\rm cp} + w_{\rm noz} -1)\; ,
  \label{eq:boss_weights}
\end{equation}
which considers the impact of seeing ($w_{\rm see}$), star
contamination ($w_{\rm star}$) and presence of a close target ($w_{\rm
  cp}$), as well as spectroscopic measurement failures ($w_{\rm
  noz}$).


\subsection{Clustering wedges}
\label{subsec:wedges}

The cosmological analysis performed in this work is based on the
redshift-space clustering wedges of the 2PCF of the spectroscopic
cluster catalog presented in \S \ref{subsec:spectroscopic}. The
clustering wedges have been introduced by \citet{kazin2010} as a
convenient projection statistic, similar to the clustering multipoles,
to compress the anisotropic 2PCF signal. The main advantage of this
approach is to reduce the dimension of the dataset to be analyzed, and
the associated covariance matrix.

To measure the three-dimensional 2PCF, we first convert the observed
coordinates of the galaxy clusters (R.A., decl., redshift) into comoving
Cartesian coordinates, assuming \citetalias{Planck2018} cosmology.

The comoving distances, $d_{\rm c}$, are related to the cosmological
redshifts, $z$, as follows:
\begin{equation}
  d_{\rm c}(z) = c \int_0^z \frac{\mbox{d}z'}{H(z')} \; ,
  \label{eq:distance}
\end{equation} 
where $c$ is the speed of light, and $H$ is the Hubble parameter,
which in a flat $\Lambda$CDM model reads as
\begin{equation}
  H(z) = H_0\left[\Omega_{\rm M}(1+z)^3+(1-\Omega_{\rm
      M})\right]^{1/2} \; .
\label{eq:Hubble}
\end{equation}
Neglecting redshift uncertainties and second-order corrections, the
observed redshift, $z_{\rm obs}$, is related to the cosmological
redshift, $z$, as follows:
\begin{equation}
z_{\rm obs} = z + \frac{v_\parallel}{c}(1+z) \; ,
\label{eq:redshift}
\end{equation}
where $v_\parallel$ is the peculiar velocity along the
line of sight. In this analysis, the impact of cluster redshift
uncertainties on the 2PCF is minor, especially on large scales, as we
consider only the spectroscopic cluster sample. Assuming that the
cluster spectroscopic redshift uncertainties follow a Gaussian
distribution \citep[e.g.][]{sereno2015}, their effects on the 2PCF are
degenerate with those of small-scale peculiar random motions, and do
not require any additional parameters to be modeled (see \S
\ref{sec:Modeling}).

Since the peculiar velocities of the analyzed cluster sample are
unknown, we estimate the comoving distances by substituting $z$ with
$z_{\rm obs}$ in Eq. \ref{eq:distance}, thus introducing the so-called
RSD. Hereafter, the redshift-space spatial coordinates are indicated
with {\em s}.

The anisotropic 2PCF in redshift-space is computed with the
\citet{landy1993} estimator:
\begin{equation}
  \hat{\xi}(s, \mu) \, = \, \frac{N_{RR}}{N_{CC}} \frac{CC(s,
    \mu)}{RR(s, \mu)} -2 \frac{N_{RR}}{N_{CR}} \frac{CR(s, \mu)}{RR(s,
    \mu)} +1 \,,
  \label{eq:xiLS}
\end{equation}
where $\mu$ is the cosine of the angle between the line of sight and
the comoving separation $s$, $CC(s, \mu)$, $RR(s, \mu),$ and $CR(s,
\mu)$ are the numbers of cluster-cluster, random-random, and
cluster-random pairs in bins of $s$ and $\mu$, i.e. in $s \pm\Delta s$
and $\mu \pm\Delta \mu$, $N_C$ and $N_R$ are the total numbers of
clusters and random objects, and $N_{CC}=N_C(N_C-1)/2$,
$N_{RR}=N_R(N_R-1)/2$, and $N_{CR}=N_CN_R$ are the total numbers of
cluster-cluster, random-random, and cluster-random pairs,
respectively. The \citet{landy1993} estimator of the 2PCF is widely
used as it provides the minimum variance when $|\xi|\ll1$, and it is
unbiased in the limit of an infinitely large random sample
\citep{keihanen2019}. We estimate the comoving separation associated
with each bin as the average cluster pair separation inside the bin
\citep[e.g.][]{zehavi2011}.

Lastly, to efficiently compress the information contained in the
clustering signal, we estimate the so-called wedges of the 2PCF
\citep{kazin2012}, that consist in the integrals of $\xi(s, \mu)$ over
wide bins of $\mu$:
\begin{equation}
  \label{eq:wedges}
  \xi_{w}(s) \equiv \frac{1}{\Delta\mu} \int_{\mu_{1}}^{\mu_{2}}
  \mathrm{d}\mu \, \xi(s, \mu) \,,
\end{equation}
where $\Delta\mu = \mu_{2}-\mu_{1}$ is the wedge width. Here, we set
$\Delta\mu=0.5$, which leads to two clustering wedges, that is the
transverse wedge, $\xi_\perp(s) \equiv \xi_{1/2}(\mu_{min} = 0, s)$,
and the radial wedge, $\xi_\parallel(s) \equiv \xi_{1/2}(\mu_{min} =
0.5, s)$, computed in the ranges of $0\leq\mu<0.5$ and
$0.5\leq\mu\leq1$, respectively.


\section{Modeling}
\label{sec:Modeling}


\subsection{Redshift-space distortions}
\label{subsec:rsd}

We model the redshift-space transverse and radial wedges of the 2PCF
of our cluster catalog with an extended version of the
\citet{taruya2010} model, which includes the nonlinear biasing model
by \citet{mcdonald2009b}. Following \citet{beutler2014}, we will refer
to this as the extended Taruya, Nishimichi $\&$ Saito (eTNS) model.

The redshift-space power spectrum of galaxy clusters in the eTNS model
is approximated as follows:
\begin{multline}
  P^{s}(k, \mu) = D(k, \mu, f, \sigma_v) \biggl[P_{\mathrm{c}, \delta
      \delta}(k) + 2 f \mu^{2} P_{\mathrm{c}, \delta \theta}(k) +
    \\ f^{2} \mu^{4} P_{\theta \theta}(k) + b_{1}^{3} C_A(k, \mu, f,
    b_1)+b_1^{4} C_B(k, \mu, f, b_1) \biggr] \, ,
  \label{eq:eTNS_model}
\end{multline}
where $f$ is the linear growth rate, $b_1$ is the linear bias,
$\theta(\mathbf{k})\equiv[-i \mathbf{k} \cdot \mathbf{v}(\mathbf{k})]
/[a f(a) H(a)]$ is the velocity divergence, $P_{\mathrm{c},
  \delta\delta}(k)$ is the real-space density cluster power spectrum,
$P_{\mathrm{c}, \delta\theta}(k)$ and $P_{\theta\theta}(k)$ are the
real-space density-velocity divergence cross-spectrum and the
real-space velocity divergence auto-spectrum of clusters,
respectively, assuming no velocity bias, i.e. $P_{\mathrm{c},
  \theta\theta}(k) = P_{ \theta\theta}(k)$, $D(k, f, \mu, \sigma_v)$
is a damping factor used to model the random peculiar motions at small
scales, and $C_A$ and $C_B$ are two additional terms to correct for
systematics at small scales. The cluster power spectra are computed
with the nonlinear biasing model by \citet{mcdonald2009b} as follows:
  \begin{multline}
    P_{\mathrm{c}, \delta \delta}(k) = b_{1}^{2} P_{\delta
      \delta}(k)+2 b_{2} b_{1} P_{b 2, \delta}(k)+2 b_{s^2} b_{1} P_{b
      s 2, \delta}(k) \\ +2 b_{3 \mathrm{nl}} b_{1} \sigma_{3}^{2}(k)
    P_{\mathrm{m}}^{\mathrm{lin}}(k)+b_{2}^{2} P_{b 22}(k) \\ +2 b_{2}
    b_{s^2} P_{b 2 s 2}(k)+b_{s^2}^{2} P_{b s 22}(k)+N \, ,
    \label{eq:Pdd}
  \end{multline}
  
  \begin{multline}
    P_{\mathrm{c}, \delta \theta}(k) = b_{1} P_{\delta
      \theta}(k)+b_{2} P_{b 2, \theta}(k)+b_{s^2} P_{b s 2,
      \theta}(k) \\ +b_{3 \mathrm{nl}} \sigma_{3}^{2}(k)
    P_{\mathrm{m}}^{\mathrm{lin}}(k)\, ,
    \label{eq:Pdt}
  \end{multline}
where $P_{\mathrm{m}}^{\mathrm{lin}}(k)$ is the linear power
spectrum. The adopted biasing model has the following four parameters,
besides the linear bias term, $b_1$: the second-order local and
nonlocal bias parameters, $b_2$ and $b_{s^2}$, the third-order
nonlocal bias parameter, $b_{3 \mathrm{nl}}$, and the constant
stochasticity term, $N$. The latter parameter affects only the
smallest comoving separations, which are not considered in our
analysis. As we verified, its impact on the cosmological outcomes of
this work is in fact negligible.  In the local Lagrangian framework
the nonlocal bias terms can be written as a function of $b_1$ as
follows \citep{chan2012, saito2014}:
\begin{equation}
  b_{s^2} = -\frac{4}{7}(b_1-1)\, ,   
\end{equation}
\begin{equation}
  b_{3 \mathrm{nl}} = \frac{32}{315}(b_1-1) \, .
\end{equation}

The $P_{\delta\delta}(k)$, $P_{\delta\theta}(k)$ and
$P_{\theta\theta}(k)$ terms are estimated in the standard perturbation
theory (SPT), which consists of expanding the statistics as a sum of
infinite terms, corresponding to the $n$-loop corrections \citep[see
  e.g.][]{gilmarin2012}. Considering corrections up to the first loop
order, the matter power spectrum can be modeled as follows:
\begin{equation}
  \label{eq:SPT_1loop}
  P^{\mathrm{SPT}}(k) = P_{\mathrm{m}}^{\mathrm{lin}}(k)+2
  P_{13}(k)+P_{22}(k) \, ,
\end{equation}
where the one-loop correction terms are computed with the \texttt{CPT
  Library}\footnote{\href{http://www2.yukawa.kyoto-u.ac.jp/~atsushi.taruya/cpt\_pack.html}
  {http://www2.yukawa.kyoto-u.ac.jp/~atsushi.taruya/cpt\_pack.html}}
\citep{taruya2008, zhao2021}. To test the model accuracy, we compared
the outcomes of our reference analysis with the ones obtained by
adopting the \citet{bel2019} universal fitting functions for
$P_{\delta\theta}(k)$ and $P_{\theta\theta}(k)$, finding negligible
differences \citep[see also][]{pezzotta2017, delatorre2017}. The other
power spectrum terms in Eqs.~\eqref{eq:Pdd} and ~\eqref{eq:Pdt}, that
is $P_{b 2, \delta}(k)$, $P_{b s 2, \delta}(k)$, $P_{b 22}(k)$, $P_{b
  2 s 2}(k)$, $P_{b s 22}(k)$, $P_{b 2, \theta}(k)$ and $P_{b s 2,
  \theta}(k)$, are computed as a function of
$P_{\mathrm{m}}^{\mathrm{lin}}(k)$, as prescribed in
e.g. \citet{beutler2014} and \citet{gilmarin2014}.

The damping term is assumed to be Lorentzian in Fourier space
\citep[see e.g.][]{delatorre2017}:
\begin{equation}
  \label{eq:streamming}
  D(k, f, \mu, \sigma_v) = \frac{1}{1+k^2f^2\mu^2\sigma_v^2} \, ,
\end{equation}
where $\sigma_v$ is a nuisance parameter to marginalize over
\citep{davis1983, fisher1994, zurek1994}.

Finally, we compute the correction terms $C_A$ and $C_B$ in
Eq.~\eqref{eq:eTNS_model} in SPT as follows
\citep{taruya2010, delatorre2012}:
\begin{align}
  \begin{split}
    \label{eq:TNS_A_term}
    C_A(k, \mu, f, b_1) &=(k \mu f) \int
    \frac{\mathrm{d}^{3}\boldsymbol{p}}{(2 \pi)^{3}}
    \frac{p_{z}}{p^{2}} \\ &
    \times\left[B_{\sigma}(\boldsymbol{p},\boldsymbol{k} -
      \boldsymbol{p},-\boldsymbol{k})-B_{\sigma}(\boldsymbol{p},
      \boldsymbol{k},-\boldsymbol{k}-\boldsymbol{p})\right]\,,
  \end{split} \\
  \label{eq:TNS_B_term}
  C_B(k, \mu, f, b_1) &=(k \mu f)^{2} \int
  \frac{\mathrm{d}^{3}\boldsymbol{p}}{(2 \pi)^{3}} F(\boldsymbol{p})
  F(\boldsymbol{k}-\boldsymbol{p}) \,, \\
  \label{eq:TNS_integral_term}
  F(\boldsymbol{p})&=\frac{p_{z}}{p^{2}}\left[P_{\delta \theta}(p)+f
    \frac{p_{z}^{2}}{p^{2}} P_{\theta \theta}(p)\right]\,,
\end{align}
where $B_\sigma$ is the cross bispectrum. The $C_A$ and $C_B$ terms
are proportional to $b_1^3$ and $b_1^4$, respectively, and can be
expressed as a power series expansion of $b_1$, $f$ and $\mu$
\citep[see e.g.][for more details]{garcia-farieta2019,
  garcia-farieta2020}.

We note that the eTNS model given by Eq.~\eqref{eq:eTNS_model} reduces
to the \citet{taruya2010} model if all the nonlinear bias terms are
neglected, to the \citet{scoccimarro2004} model if also the $C_A$ and
$C_B$ terms are neglected, and to the so-called dispersion model if
both $P_{\delta\theta}(k)$ and $P_{\theta\theta}(k)$ are approximated
as $P_{\delta\delta}(k)$, which is valid in the linear regime
\citep{kaiser1987, peacock1996}.

\begin{figure*}
  \includegraphics[width=0.49\textwidth]{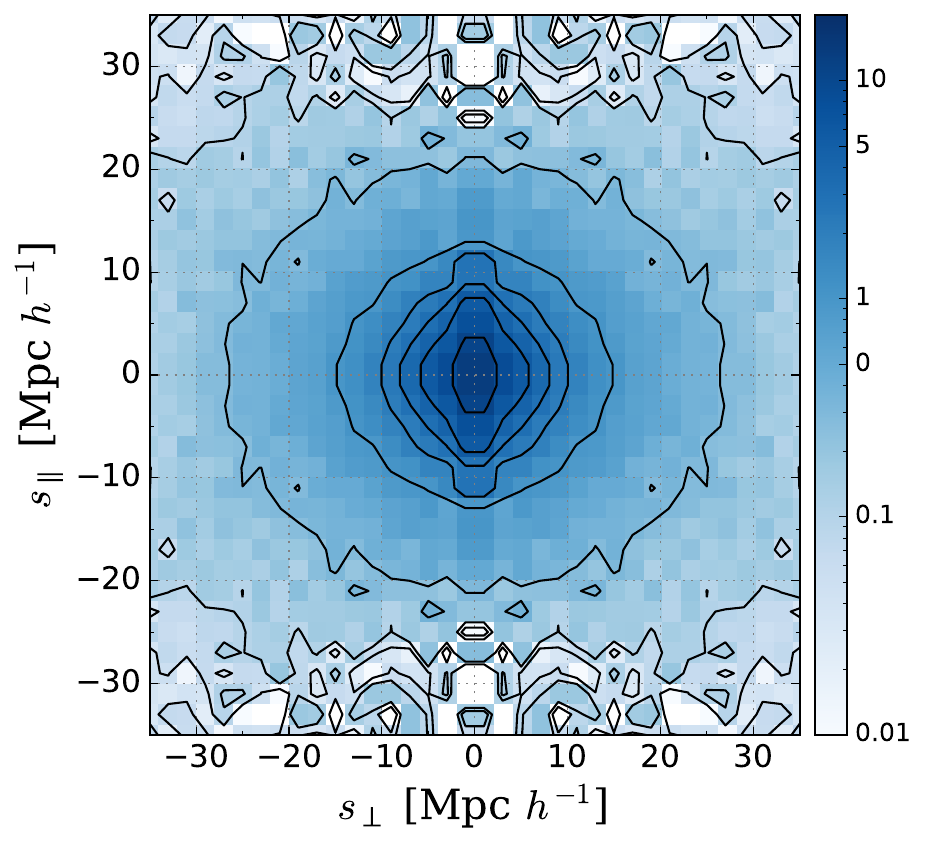}
  \includegraphics[width=0.49\textwidth]{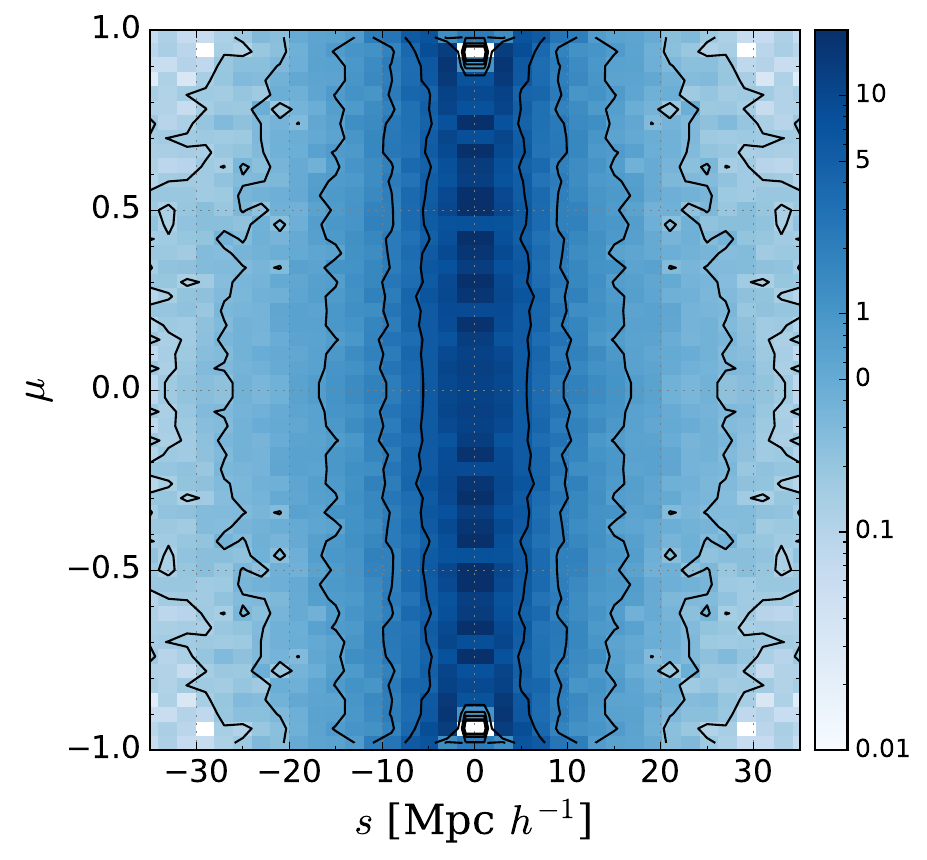}
  \caption{The redshift-space 2PCF of the spectroscopic cluster
    catalog. {\em Left panel}: 2PCF in Cartesian coordinates, which is
    as a function of perpendicular and parallel separations to the
    line of sight. {\em Right panel}: 2PCF in polar coordinates, that
    is as a function of distance modulus and cosine of the angle
    between the line of sight.}
  \label{fig:2D}
\end{figure*}

The power spectrum multipoles can be estimated from
Eq.~\eqref{eq:eTNS_model}, as follows:
  \begin{equation}
    P_l(k) =
    \frac{2l+1}{2\alpha_\perp^2\alpha_\parallel}\int_{-1}^1\mbox{d}
    \mu\, P^s(k^\prime, \mu^\prime)L_l(\mu) \, ,
    \label{eq:Pkmult_model}
  \end{equation}
where the \citet{alcock1979} (AP) geometric distortions, caused by a
possibly incorrect assumption of the background cosmology used to
convert cluster redshifts into comoving distances in
Eq.~\eqref{eq:distance}, are modeled by rescaling the wave numbers as
follows \citep{beutler2014}:

\begin{equation}
  k^\prime = \frac{k}{\alpha_\perp} \left[1 + \mu^2 \left(
    \frac{\alpha^{2}_\perp}{\alpha^{2}_\parallel} -1
    \right)\right]^{1/2} \, ,
\end{equation}
\begin{equation}
  \mu^\prime = \mu\frac{\alpha_\perp}{\alpha_\parallel} \left[1 +
    \mu^2 \left( \frac{\alpha^{2}_\perp}{\alpha^{2}_\parallel} - 1
    \right)\right]^{-1/2}\, ,
\end{equation}

with

\begin{equation}
  \alpha_\parallel = \frac{H^{\rm fid}(z)r_s^{\rm
      fid}(z_d)}{H(z)r_s(z_d)} \, ,
\end{equation}

\begin{equation}
  \alpha_\perp = \frac{D_A(z)r_s^{\rm fid}(z_d)}{D_A^{\rm
      fid}(z)r_s(z_d)} \, ,
\end{equation}
where $H^{\rm fid}(z)$ and $D_A^{\rm fid}(z)$ are the fiducial values
for the Hubble constant and angular diameter distance, respectively,
and $r_s^{\rm fid}(z_d)$ is the fiducial sound horizon at the drag
redshift assumed in the power spectrum template.

The corresponding 2PCF multipoles in configuration space read as
\begin{equation}
  \xi_l(s) = i^l\int_{-\infty}^{\infty}\frac{\mbox{d}
    k}{2\pi^2}k^2P_l(k)j_l(ks) \, , 
  \label{eq:mult_model}
\end{equation}
where $L_l(\mu)$ are the Legendre polynomials and $j_l$ are the
spherical Bessel functions of order $l$.

Lastly, we assess the redshift-space wedges from the multipole moments
through the following relation:
\begin{equation}
  \xi_{w}(r)=\sum_l\xi_l(s)\bar{L}_l \, ,
  \label{eq:wedges_multipoles}
\end{equation}
where $\bar{L}_l$ is the average value of the Legendre polynomials
over the interval $[\mu_1,\mu_2]$. In particular, neglecting minor
contributions from multipoles with $l>2$ and considering the wedge
width $\Delta\mu=0.5$, Eq.~\eqref{eq:wedges_multipoles} can be written
as follows \citep{kazin2012}:
\begin{equation}
  \left(\begin{array}{l}{\xi_{ \|}}
    \\ {\xi_{\perp}}\end{array}\right)=\left(\begin{array}{cc}{1} &
    {\frac{3}{8}} \\ {1} &
    {-\frac{3}{8}}\end{array}\right)\left(\begin{array}{c}{\xi_{0}}
    \\ {\xi_{2}}\end{array}\right) \, .
  \label{eq:wedges_multipoles2}
\end{equation}

A full validation of the implemented likelihood algorithms on
simulated galaxy and cluster catalogs will be presented in a
forthcoming paper \citep{garcia-farieta2021}.

As a common practice, and to directly compare to previous similar
analyses performed on galaxy samples, we parameterize the model as a
function of $\sigma_v$ and the three parameter products
$\left[f\sigma_8,\, b_1\sigma_8,\, b_2\sigma_8\right]$ \citep[but
  see][]{sanchez2020}, fixing the other parameters to
\citetalias{Planck2018} cosmology.

In this work we focus on scales smaller than the BAO ones, which does
not allow us to put strong enough constraints on the geometric
distortions that are degenerate with RSD
\citep[e.g.][]{taruya2011}. Nevertheless, to marginalize over the AP
distortion parameters, we allow them to vary, considering Gaussian
priors with standard deviation of $0.01$ \citep[see e.g.][for a
  similar approach]{delatorre2017}.

The posterior distribution constraints on these parameters are
assessed through a Markov Chain Monte Carlo (MCMC) statistical
analysis, assuming a standard Gaussian likelihood:
\begin{equation}
  -2 \ln \mathcal{L}=\sum_{i, j=1}^{N}\left[\xi_{k}^{D}(s_{i}) -
    \xi_{k}^{M}(s_{i})\right] C_{k}^{-1}(s_{i},
  s_{j})\left[\xi_{k}^{D}(s_{j}) - \xi_{k}^{M}(s_{j})\right]\,,
\end{equation}
where $N$ is the number of bins at which the wedges are computed, and
the superscripts $D$ and $M$ refer to data and model, respectively.

The covariance matrix, $C_{k}$, which measures the variance and
correlation between 2PCF wedge bins, is defined as follows:
\begin{equation}
  C_{w}\left(s_{i}, s_{j}\right)=\frac{1}{N_{R}-1}
  \sum_{n=1}^{N_{R}}\left[\xi_{w}^{n}\left(s_{i}\right)-\overline{\xi}_{w}
    \left(s_{i}\right)\right]\left[\xi_{w}^{n}
    \left(s_{j}\right)-\overline{\xi}_{w}\left(s_{j}\right)\right] \,
  .
\end{equation}
The indices $i$ and $j$ run over the 2PCF wedge bins, while $w=0,0.5$
refers to each clustering wedge. In both cases,
$\bar{\xi}_w=1/N_R\sum_{n=1}^{N_R}\xi_w^n$ is the average wedge of the
2PCF, and $N_R=100$ is the number of realizations obtained by
resampling the catalogs with the bootstrap method. We correct the
inverse covariance matrix estimator to account for the finite number
of realizations as in \citet{hartlap2007}, while the parameter
uncertainties are corrected to take into account the uncertainties in
the covariance estimate as in \citet{percival2014}.


\subsection{Exploiting cluster masses}
\label{subsec:cluster_bias}

Similarly to RSD analyses of galaxy clustering, we adopt large flat
priors on $f\sigma_8$, $b_2\sigma_8$ and $\sigma_v$, specifically
$f\sigma_8=\left[0, 1\right]$, $b_2\sigma_8=\left[-10, 10\right]$ and
$\sigma_v=\left[0, 100\right] \mbox{Mpc}\,h^{-1}$, respectively. While
$\sigma_8$ could be constrained directly from the cluster mass
function, this would have required an accurate knowledge of the
cluster selection function to avoid systematic uncertainties. To
provide conservative linear growth constraints, we prefer to focus the
current analysis on cluster clustering, setting all the cosmological
parameters, including $\sigma_8$, to \citetalias{Planck2018}
values. The constraint on the linear growth rate we will derive in
this paper has to be considered in this respect, though to compare to
previous analyses we will express our results in terms of $f\sigma_8$.

Differently from galaxy clustering analyses, we can set a strong prior
on $b_1\sigma_8$ thanks to the knowledge of galaxy cluster masses
inferred from weak-lensing scaling relations. In particular, the
$b_1\sigma_8$ prior is centered on the effective linear bias of the
cluster sample, which is estimated as in \citet{marulli2018}:
\begin{equation}
  b_{\rm eff}^2 = <b(\tilde{M}_i,z_i)b(\tilde{M}_j,z_j)> \; ,
\label{eq:bias_eff}
\end{equation}
where the linear bias of each cluster, $b$, is computed with the
\citet{tinker2010} model, while $\tilde{M_i}$ and $\tilde{M_j}$ are
the masses of the two clusters of each pair, at redshifts $z_i$ and
$z_j$, respectively, estimated from the weak-lensing cluster mass
scaling relation given by Eq.~\eqref{eq:scaling}.  This represents the
key difference with respect to analogous RSD analyses of galaxy
samples, as in those cases the masses of the dark matter haloes
hosting the galaxies are unknown and no priors can be reliably assumed
on the bias of the catalog. We will discuss the impact of this
assumption in \S \ref{sec:Results}.

Drawing a set of mass samples from the scaling relation, we computed
the average bias and variance of our cluster catalog, which correspond
to $b_1\sigma_8=1.722\pm0.002$. We consider a Gaussian prior on
$b_1\sigma_8$ centered on the latter value. To provide conservative
constraints, we adopt a prior width $5$ times larger than the
estimated standard deviation, i.e., $0.01$, to include possible
systematic uncertainties in the adopted bias model and scaling
relation.


\section{Results}
\label{sec:Results}


\subsection{Constraints on the growth rate}
\label{subsec:constraints}

\begin{figure}
  \includegraphics[width=0.48\textwidth]{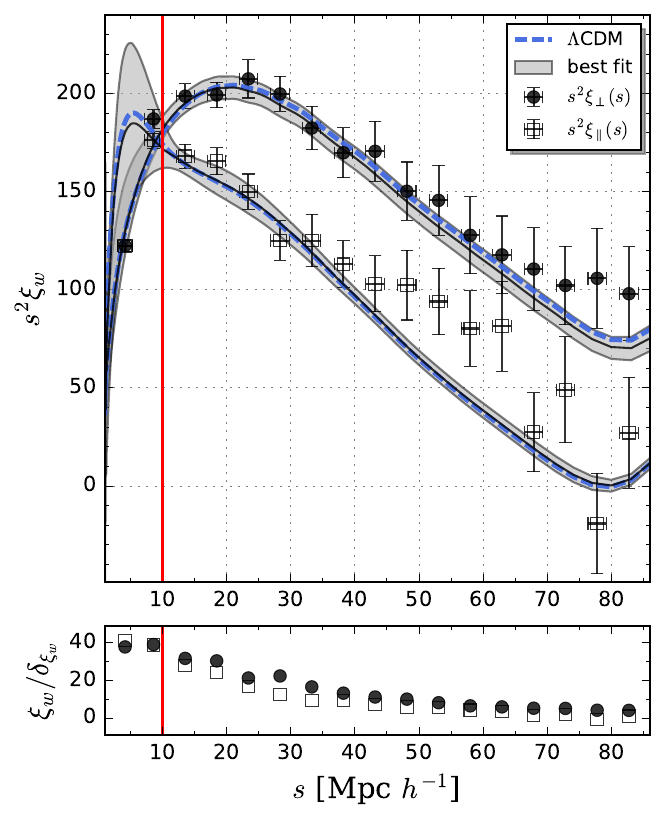}
  \caption{{\em Upper panel}: The redshift-space transverse (solid
    circles) and radial (open squares) wedges of the spectroscopic
    cluster catalog compared to the best-fit eTNS model, which is the
    median of the MCMC posterior distribution (black solid line). The
    shaded areas show the $68\%$ uncertainty on the posterior
    median. The horizontal error bars are the standard deviation
    around the mean pair separation in each bin. The vertical error
    bars show the diagonal values of the bootstrap covariance
    matrix. The vertical red line indicates the minimum scale used in
    the fitting analysis. The minor mismatch in the radial wedge at
    $40<s\left[\mbox{Mpc}\,h^{-1}\right]<60$ is not statistically
    significant considering the covariance in the measurements
    ($\tilde{\chi}^2=0.71$). {\em Lower panel}: The signal-to-noise
    ratio, that is the transverse (solid circles) and radial (open
    squares) wedge values divided by the corresponding standard
    deviations.}
  \label{fig:wedges}
\end{figure}

\begin{figure}
  \includegraphics[width=0.49\textwidth]{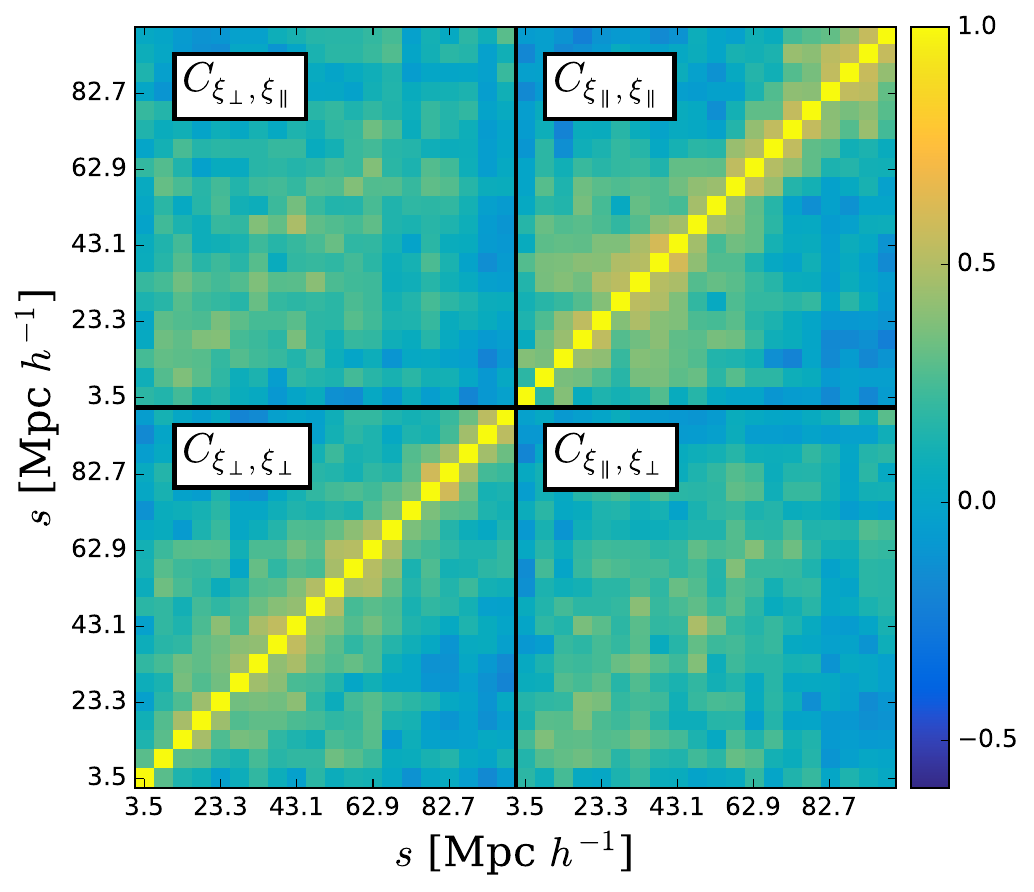}
  \caption{The bootstrap correlation matrix
    $\left(C_{i,j}/\sqrt{C_{i,j}C_{j,i}}\right)$ of the redshift-space
    transverse and radial wedges of the spectroscopic cluster
    catalog.}
  \label{fig:corr}
\end{figure}

As described in \S \ref{sec:ClusteringMeasurements}, the clustering
wedges are computed by integrating the redshift-space 2PCF, $\xi(s,
\mu)$, over two bins of $\mu$.  In Fig. \ref{fig:2D} we present the
redshift-space 2PCF of the selected cluster sample in two coordinate
systems, i.e., as a function of perpendicular ($s_\perp$) and parallel
($s_\parallel$) separations to the line of sight (Cartesian
coordinates), and as a function of distance modulus ($s$) and cosine
of the angle ($\mu$) between the line of sight (polar coordinates).
In real space the contour lines of the former statistics would be
circular, while the ones of the latter statistics would be
straight. RSD introduce anisotropies in the derived map that warp
these contour lines, an effect that depends directly on the value of
the linear growth rate of cosmic structures.

The shape of the Cartesian 2PCF of the selected clusters shown in the
left panel of Fig. \ref{fig:2D} appears similar to the one of
galaxies, as expected \citep[e.g.][]{alam2017, marulli2017}. In fact,
as described in \S \ref{subsec:spectroscopic}, the 2PCF of galaxy
clusters we measure in this work coincides with the 2PCF of BCGs, by
construction. Nevertheless, the cosmological analysis of this dataset
provides a clear advantage as, differently from galaxy clustering
studies, we can infer in this case the linear bias of the sample from
the richness-mass scaling relation of the galaxy clusters hosting the
selected BCGs, as explained in \S \ref{subsec:cluster_bias}.

The Fingers-of-God distortions at small scales due to incoherent
peculiar motions are not completely negligible, though much less
strong relative to the case of the 2PCF of lower biased tracers
\citep[see the discussion in][]{marulli2017}. The polar 2PCF shown in
the right panel of Fig. \ref{fig:2D} is the statistics we integrate
along the $\mu$ direction to compute the wedges.

Figure \ref{fig:wedges} shows the redshift-space transverse and radial
wedges of the cluster 2PCF, defined by Eq.~\eqref{eq:wedges}. The
horizontal and vertical error bars are the standard deviation around
the mean pair separation in each bin, and the diagonal values of the
bootstrap covariance matrix, respectively. In real space, the radial
and transverse wedges would be statistically equal, for isotropy. On
the other hand, redshift-space anisotropies make these two statistics
significantly different, as shown in Fig. \ref{fig:wedges}.

\begin{figure}
  \includegraphics[width=0.46\textwidth]{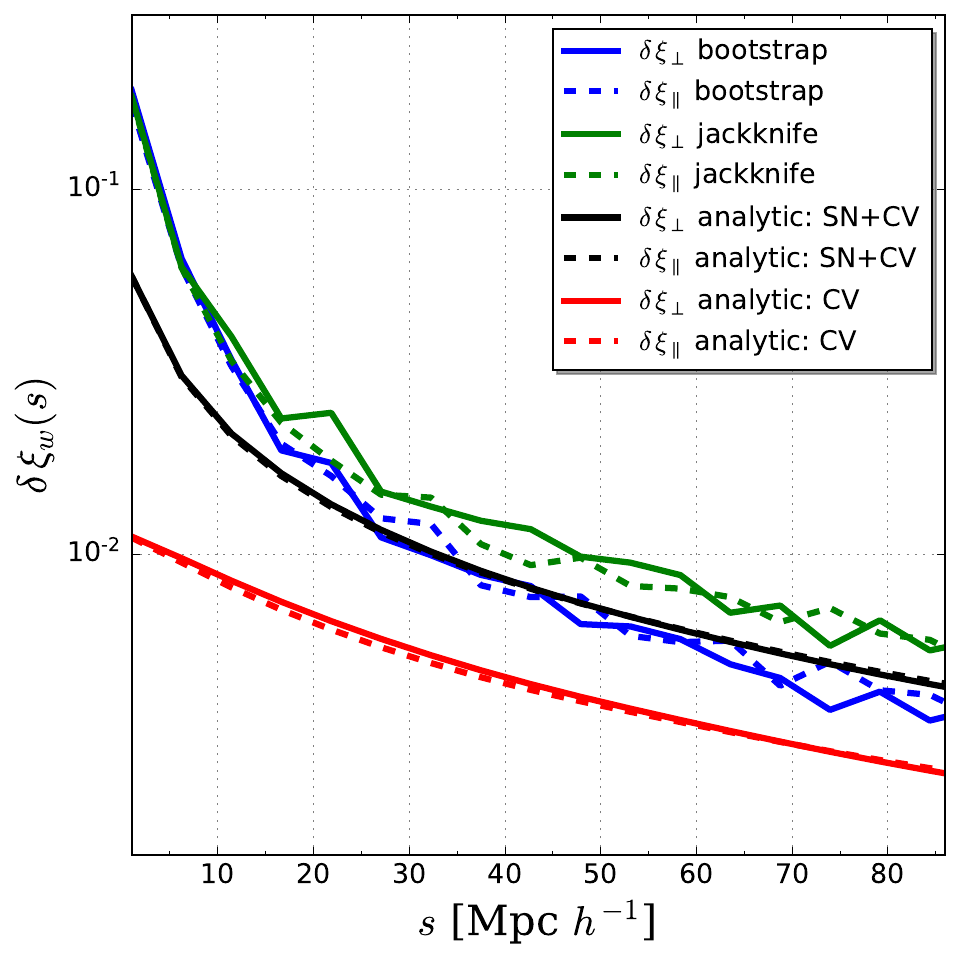}
  \caption{Comparison between the diagonal values of different
    covariance matrix estimates of the redshift-space transverse
    (solid lines) and radial (dashed lines) wedges of the
    spectroscopic cluster catalog. Blue and green lines refer to the
    bootstrap and jackknife covariance matrix, respectively. Black
    lines refer to the theoretical covariance matrix by
    \citet{grieb2016}, comprising both the shot noise (SN) and cosmic
    variance (CV) contributions, while red lines include only the
    contribution of cosmic variance.}
  \label{fig:errors}
\end{figure}

The correlation matrix, which is $C_{i,j}/\sqrt{C_{i,j}C_{j,i}}$, of
the redshift-space transverse and radial wedges is shown in
Fig. \ref{fig:corr}. The algorithms to estimate the covariance matrix
have been highly validated in previous works on both simulations and
real cluster catalogs \citep[e.g.][]{veropalumbo2014, veropalumbo2016,
  marulli2017, garcia-farieta2020}. To further check the results of
the current analysis, we compare the bootstrap error estimates with
the ones obtained with either the jackknife method or the analytic
Gaussian model provided by \citet{grieb2016} \citep[for the
  theoretical modeling of non-Gaussian contributions to the
  covariance, which are neglected in the current analysis,
  see][]{sugiyama2020}. The diagonal values of the bootstrap,
jackknife, and analytic matrices are compared in
Fig. \ref{fig:errors}. The estimated bootstrap clustering
uncertainties are statistically consistent with the analytic ones at
scales larger than about $20$ \Mpch, while the jackknife uncertainties
appear slightly larger.

The bootstrap method allows us to draw a greater number of
realizations relative to the jackknife method, providing a smoother
covariance matrix, whose inverse is less affected by numerical
noise. Moreover, it does not depend on free parameters, differently
from the analytic covariance matrix, which depends on the sample bias
and on the effective area of the survey, whose values are inferred
within uncertainties. For the above reasons, in this work we rely on
the bootstrap covariance uncertainties.

Following the method described in \S \ref{sec:Modeling}, we perform a
joint statistical analysis of the redshift-space radial and transverse
wedges of the selected spectroscopic cluster catalog, in the
standard $\Lambda$CDM framework. The best-fit eTNS model obtained from
the median of the MCMC posterior distribution is reported in
Fig. \ref{fig:wedges}, together with its $68\%$ uncertainty
region. The fit is performed in the comoving scale range
$10<s\left[\mbox{Mpc}\,h^{-1}\right]<80$. The model appears
statistically consistent with the measurements in the scale range
considered. We note only a minor mismatch in the radial wedge at
$40<s\left[\mbox{Mpc}\,h^{-1}\right]<60$, though it is not
statistically significant considering the covariance in the
measurements. The reduced $\chi^2$ at the best-fit eTNS model is
$\tilde{\chi}^2=-2 \ln \mathcal{L}/\mbox{d.o.f.}=0.71$ (where
$\mbox{d.o.f.}$ are the degrees of freedom of the data sample).

\begin{figure}
  \includegraphics[width=0.49\textwidth]{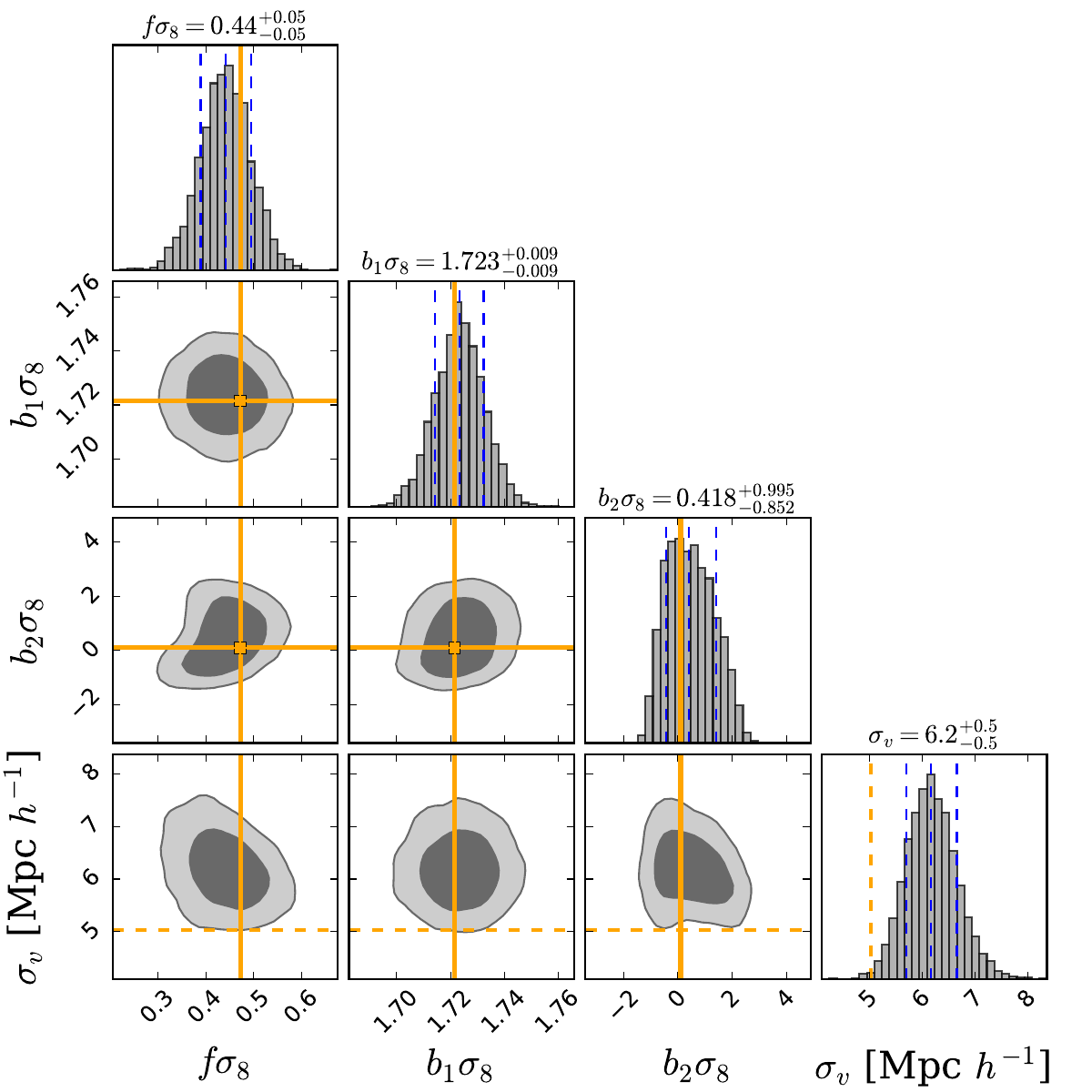}
  \caption{$68\%$ and $95\%$ posterior confidence regions for the four
    eTNS parameters $f\sigma_8$, $b_1\sigma_8$, $b_2\sigma_8$,
    $\sigma_v$, obtained from the MCMC modeling of the redshift-space
    radial and transverse wedges of galaxy clusters in the scale range
    $10<s\left[\mbox{Mpc}\,h^{-1}\right]<80$. The vertical dashed blue
    lines show the first quantiles of the 1D marginalized
    distributions. The solid orange lines show the
    \citetalias{Planck2018} $f\sigma_8$, $b_1\sigma_8$ and
    $b_2\sigma_8$ predictions, with $b_1$ being estimated by
    Eq.~\eqref{eq:bias_eff} with \citet{tinker2010} model, and $b_2$
    from Eq.~\eqref{eq:b2}. The dashed orange lines show the
    linear-order estimate of the one-dimensional velocity dispersion
    given by Eq.~\eqref{eq:sv}.}
  \label{fig:constraints}
\end{figure}

The marginalized posterior distributions on $f\sigma_8$,
$b_1\sigma_8$, $b_2\sigma_8$ and $\sigma_v$ are reported in
Fig. \ref{fig:constraints}, together with the $68\%$ and $95\%$
posterior confidence regions. We assess the best-fit values and
marginalized constraints from the median and percentile values of the
posterior distribution. We note in particular that the $f\sigma_8$
posterior is consistent with a Gaussian distribution. We get
$f\sigma_{8}=0.44\pm0.05$, at the mean pair redshift $z=0.275$. The
relative statistical uncertainty is about $10\%$, which is a
remarkable result considering the sparsity of the spectroscopic
cluster sample considered. This is caused by the narrow Gaussian prior
on the effective bias of the sample, assessed through the cluster
mass-richness scaling relation, as described in \S
\ref{sec:Modeling}. In fact, the posterior distribution we get on
$b_1\sigma_8$ is statistically indistinguishable from the assumed
Gaussian prior distribution.

We tested the impact of this assumption by running the statistical
analysis assuming prior standard deviations of $0.002$, $0.05$ and
$0.1$. In particular, the latter prior width accounts for possible
bias model uncertainties of about $5\%$. We obtained
$f\sigma_{8}=0.44\pm0.05$, $f\sigma_{8}=0.42\pm0.06$ and
$f\sigma_{8}=0.41\pm0.06$, respectively, which are all statistically
consistent, considering current measurement uncertainties. Assuming
instead a flat prior distribution on $b_1\sigma_8$ in $\left[0.1,
  5\right]$, we get $f\sigma_{8}=0.40\pm0.07$, which is still
statistically consistent with our fiducial result, though with a
larger relative error of about $18\%$.

All the results obtained for the different prior assumptions
considered are statistically consistent, showing that the accuracy in
the mass estimates and in the linear bias model is high enough for the
current analysis. Improving the mass and bias modeling will become
crucial instead for cluster clustering analyses of next-generation
surveys.

The reference value of the quadratic bias factor, $b_2\sigma_8$,
reported in Fig. \ref{fig:constraints} is computed with the polynomial
relation provided by \citet{lazeyras2016}:
\begin{equation}
  b_2(b_1) = 0.412 - 2.143\,b_1 + 0.929\,b_1^2 + 0.008\,b_1^3 \; ,
  \label{eq:b2}
\end{equation}
while the reference value of $\sigma_v$ is estimated in linear
theory as follows \citep{taruya2010}:
\begin{equation}
  \sigma_{\mathrm{v}, \mathrm{lin}}^{2} = \frac{1}{3} \int
  \frac{\mathrm{d}^{3} \boldsymbol{q}}{(2 \pi)^{3}}
  \frac{P_{\mathrm{m}}^{\mathrm{lin}}(q, z)}{q^{2}} \; .
  \label{eq:sv}
\end{equation}

To test the robustness of our results, we repeated the analysis
fitting the wedges in narrower scale ranges, considering either
jackknife or analytic clustering uncertainties instead of boostrap,
and modeling small-scale random motions with a Gaussian damping term
instead of a Lorentzian one \citep{marulli2012b, sridhar2017,
  garcia-farieta2020}. Overall we found consistent results, within the
$68\%$ confidence region.

As discussed in \S \ref{subsec:rsd}, we considered tight Gaussian
priors on the AP distortion parameters with standard deviation of
$0.01$. To investigate the impact of this assumption, we run our
analysis also for standard deviation values of $0$, $0.05$ and $0.1$,
obtaining $f\sigma_{8}=0.44\pm0.05$, $f\sigma_{8}=0.4\pm0.1$ and
$f\sigma_{8}=0.3\pm0.2$, respectively. As expected, the impact of this
prior is significant. Since the current analysis focuses on scales
below the BAO peak, we cannot break the degeneracy between RSD and
geometric distortions. Constraints from a joint RSD+BAO analysis of
both the two-point and three-point correlation functions of this
cluster sample is presented in \citet{veropalumbo2021}.

\begin{figure*}
  \includegraphics[width=1\textwidth]{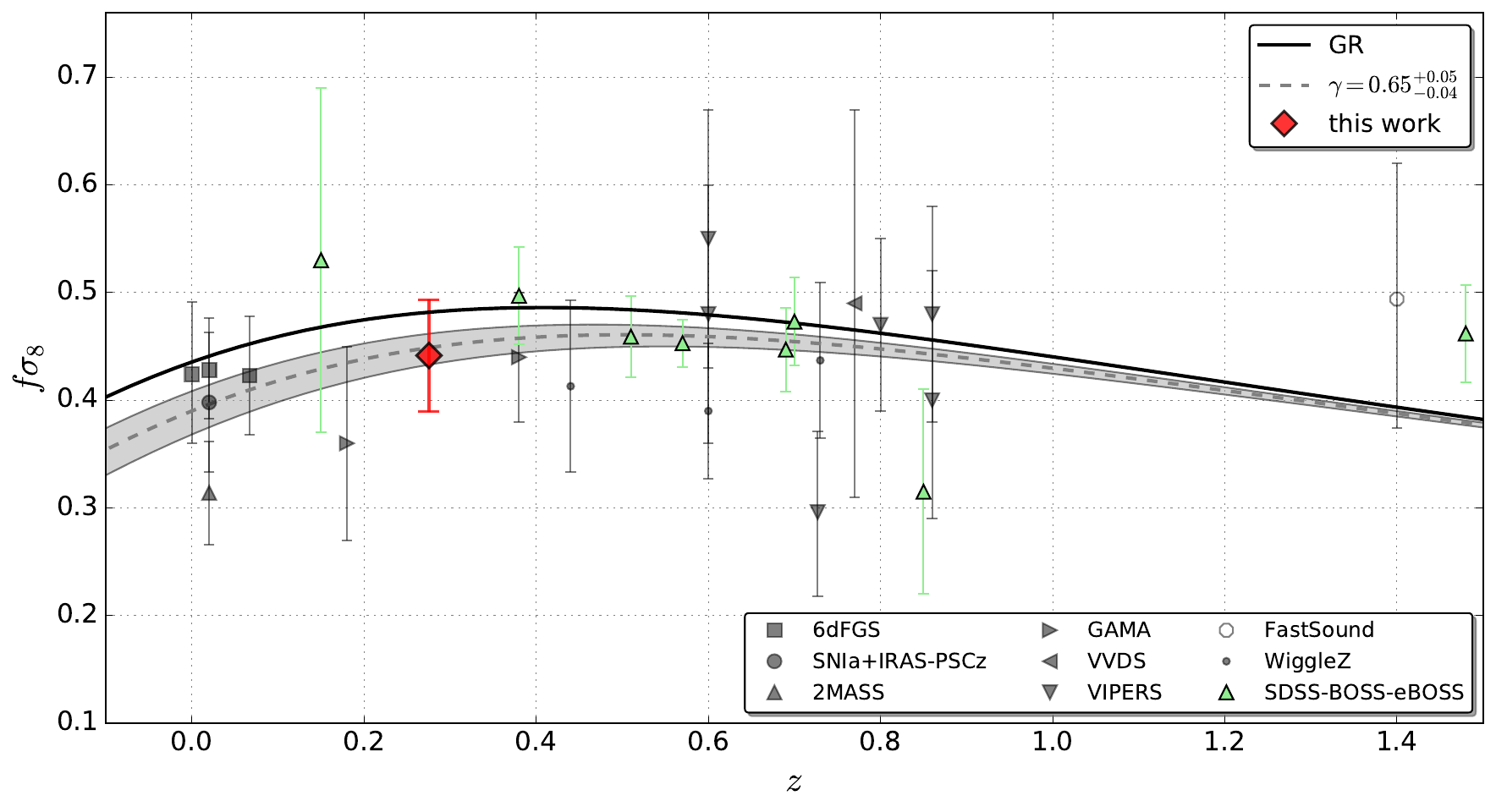}
  \caption{Constraints on $f\sigma_8$ from the redshift-space wedges
    of the galaxy cluster sample considered in this work (red diamond)
    compared to a compilation of recent measurements exploiting
    different techniques applied to the following surveys: 6dFGS
    \citep{beutler2012, huterer2017, adams2017}; `First Amendment' set
    of SNe peculiar velocities \citep{turnbull2012}; 2MASS
    \citep{davis2011}; GAMA \citep{blake2013}; WiggleZ
    \citep{blake2012}; VVDS \citep{guzzo2008}; VIPERS
    \citep{delatorre2013b, delatorre2017, pezzotta2017, hawken2017,
      mohammad2018}; FastSound \citep{okumura2016}; SDSS+BOSS+eBOSS
    \citep{eBOSS2021}. The black solid line shows the standard
    $\Lambda$CDM + GR \citetalias{Planck2018} prediction, while the
    dashed gray line and shaded area show the model with
    $\gamma=0.65^{+0.05}_{-0.04}$ \citep{moresco2017}.}
  \label{fig:fsigma8}
\end{figure*}


\subsection{Comparison to previous data and models}
\label{subsec:comparison}

In Fig. \ref{fig:fsigma8} we compare the $f\sigma_8$ constraint
obtained in this work with a large collection of measurements at
different redshifts from galaxy, quasar, and cosmic void samples and
other tracers. The data shown provide the key observables to test the
gravity theory on the largest cosmological scales\footnote{The table
  containing the $f\sigma_8$ values shown in the figure is available
  at:
  \href{https://gitlab.com/federicomarulli/CosmoBolognaLib/tree/master/External/Data/}{gitlab.com/federicomarulli/CosmoBolognaLib/External/Data}
  .}. Figure \ref{fig:fsigma8} summarizes our current understanding of
the cosmological evolution of the linear growth rate of cosmic
structures. Specifically, the data reported are from 6dF Galaxy Survey
(6dFGS) \citep{beutler2012, huterer2017, adams2017}; `First Amendment'
set of SNe peculiar velocities \citep{turnbull2012}; 2MASS
\citep{davis2011}; Galaxy And Mass Assembly (GAMA) \citep{blake2013};
WiggleZ \citep{blake2012}; VIMOS-VLT Deep Survey (VVDS)
\citep{guzzo2008}; VIMOS Public Extragalactic Redshift Survey (VIPERS)
\citep{delatorre2013b, delatorre2017, pezzotta2017, hawken2017,
  mohammad2018}; FastSound \citep{okumura2016}; SDSS+BOSS+eBOSS
\citep{howlett2015, alam2017, nadathur2019, nadathur2020,
  gilmarin2020, tamone2020, neveux2020, eBOSS2021, hou2021,
  demattia2021, bautista2021}. The latest, most constraining
measurements from SDSS+BOSS+eBOSS are highlighted in green.

The data are compared to the standard $\Lambda$CDM + general
relativity (GR) predictions, which is
$f\sigma_8=\Omega_M(z)^{0.545}\sigma_8(z)$, where $\sigma_8$ and
$\Omega_M$ are computed assuming the \citetalias{Planck2018}
cosmological parameters. The $f\sigma_8$ constraint obtained in this
work appears fully consistent with the other data, with a competitive
statistical uncertainty. By comparison, we also plot the model with
$\gamma=0.65^{+0.05}_{-0.04}$, which provides a better fit to the
$H(z)/H_0-f(z)\sigma_8(z)$ diagram, as found by \citet{moresco2017}.

\begin{figure}
  \includegraphics[width=0.48\textwidth]{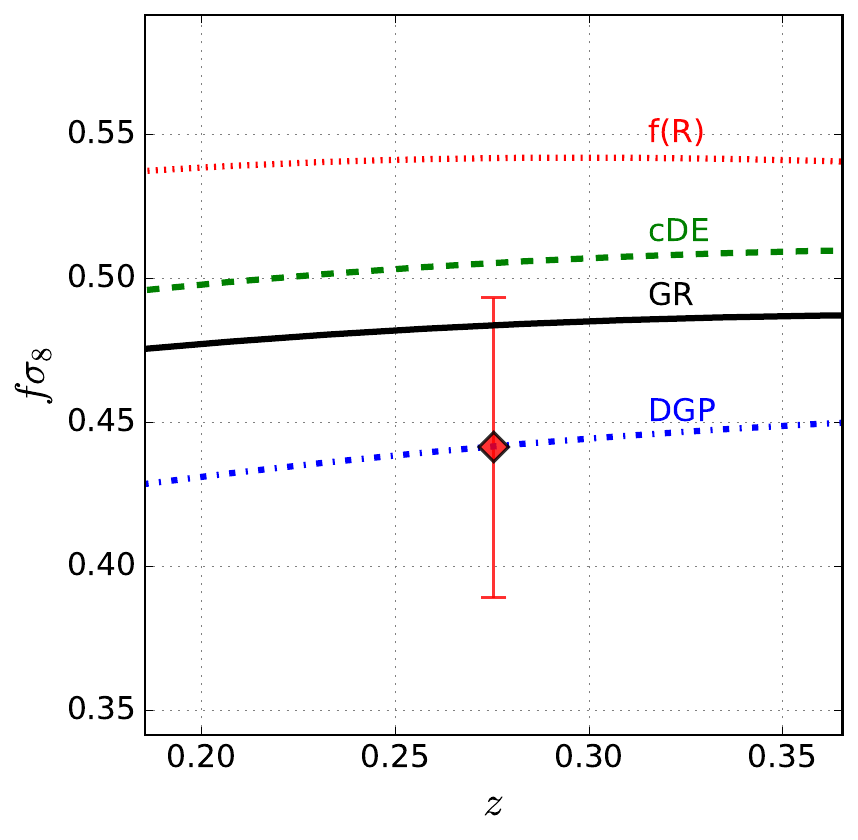}
  \caption{Comparison between the $f\sigma_8$ constraint obtained in
    this work (red diamonds) with the predictions of different gravity
    models: GR (black solid line), f($R$) (red dotted line), cDE
    (green dashed line), DGP (blue dash-dotted line).}
  \label{fig:fsigma8_MG}
\end{figure}

The exploitation of growth rate measurements at different redshifts to
discriminate among alternative dark energy and modified gravity models
would require a detailed study which is outside the scope of this
work. Nevertheless, to highlight the constraining power of current
cluster clustering measurements, we compare in
Fig. \ref{fig:fsigma8_MG} the $f\sigma_8$ constraint obtained in this
work with the predictions of three popular alternative models, that is
the f($R$) model \citep[e.g.][]{defelice2010}, the coupled dark energy
(cDE) model \citep{wetterich1995, amendola2000} and the
Dvali-Gabadaze-Porrati (DGP) model \citep{Dvali2000}.

The linear growth rate in f($R$) and DGP models can be expressed with
the so-called $\gamma$-parameterization:
\begin{equation}
  f=\Omega_M(z)^{\gamma(z)} \, ,
\end{equation}
where 
\begin{equation}
  \gamma(z)=\gamma_0+\gamma_1\frac{z}{1+z} \, .
\end{equation}
In particular, we consider the f($R$) model by \citet{hu2007}, setting
the two free model parameters to $n=2$ and $\lambda=3$ following
\citet{diporto2012}, which corresponds to $\gamma_1=0.43$ and
$\gamma_2=-0.2$ in the limit of small, still linear, scales
\citep{gannouji2009}. For the DGP model we consider the flat-space
case in which $\gamma_1=0.633$ and $\gamma_2=0.041$
\citep{maartens2006, fu2009}. Finally, we model the linear growth rate
in the cDE model with the so-called $\eta$-parameterization
\citep{diporto2008}:
\begin{equation}
  f=\Omega_M(z)^{\gamma}(1+\eta) \, ,
\end{equation}
where $\eta=c\beta_c^2$ quantifies the coupling strength.  In
particular, we consider the case with $\gamma=0.56$, $c=2.1$ and
$\beta_c=0.16$, which implies $\eta=0.056$ \citep{diporto2008,
  diporto2012}.

As shown in Fig. \ref{fig:fsigma8_MG}, the current measurement
uncertainties are not low enough to discriminate between these
alternative cosmological scenarios at sufficient statistical
level. Next-generation experiments will have instead the required
accuracy to achieve this key scientific task
\citep[e.g.][]{amendola2018}.

\section{Conclusions}
\label{sec:Conclusions}

In this work we provided new constraints on the linear growth rate of
cosmic structures from the redshift-space 2PCF of a large
spectroscopic cluster sample extracted for the BOSS survey. Cluster
clustering is a novel cosmological probe that can now be fully
exploited thanks to the large cluster samples currently available,
providing cosmological constraints complementary to the ones from
standard galaxy clustering RSD analyses. The main advantage of this
probe relative to galaxy clustering is the possibility to estimate
cluster masses. In particular, taking advantage of the information
coming from the weak-lensing cluster mass-richness scaling relation,
we could set a sharp prior on the effective bias of the sample.

The main results of this work can be summarized as follows:

\begin{itemize}
\item
We constructed a large spectroscopic catalog of optically selected
clusters from the SDSS in the redshift range $0.1<z<0.42$. The
selected sample consists of $43743$ clusters, whose angular
coordinates and redshifts are defined as the ones of their BCGs. The
cluster masses have been estimated from weak-lensing calibrated
scaling relations.

\item
  We measured the redshift-space 2PCF, as well as the transverse and
  radial 2PCF wedges, finding results consistent with theoretical
  expectations.

\item
  Assuming a $\Lambda$CDM cosmological model with
  \citetalias{Planck2018} parameters, we modeled the 2PCF wedges with
  the eTNS model. We performed a MCMC Bayesian analysis to sample the
  posterior distribution of $f\sigma_8$, $b_1\sigma_8$, $b_2\sigma_8$,
  $\sigma_v$. The cluster masses are used to set a robust prior on
  $b_1\sigma_8$.

\item
  We get $f\sigma_{8}=0.44\pm0.05$ at the mean pair redshift
  $z=0.275$, which is fully consistent with $\Lambda$CDM + GR
  predictions, and with a statistical uncertainty that is competitive
  with the current state-of-the-art constraints from other probes.

\end{itemize}

Next-generation projects like the extended Roentgen Survey with an
Imaging Telescope Array (eROSITA) satellite
mission\footnote{\href{http://www.mpe.mpg.de/eROSITA}{http://www.mpe.mpg.de/eROSITA}}
\citep{merloni2012}, the NASA's {\em Nancy Grace Roman} Space
Telescope\footnote{\href{https://nasa.gov/wfirst}{https://nasa.gov/wfirst}}
\citep{spergel2015}, the ESA {\em Euclid}
mission\footnote{\href{http://www.euclid-ec.org}{http://www.euclid-ec.org}}
\citep{laureijs2011, sartoris2016, amendola2018} and the Vera C. Rubin
Observatory LSST\footnote{Legacy Survey of Space and Time;
  \href{http://www.lsst.org}{http://www.lsst.org}} \citep{LSST2012}
will provide huge well-characterized cluster samples up to high
redshifts. While the main cosmological probe to be exploited is the
redshift evolution of cluster number counts, this work also
demonstrates that the clustering of galaxy clusters provides key
cosmological information. In particular, robust constraints on the
cosmic growth rate can be extracted from the redshift-space
anisotropies, thus testing the gravity theory on cosmological
scales. Moreover, BAO in cluster clustering provide a powerful
independent cosmological probe \citep{veropalumbo2014,
  veropalumbo2016}. In \citet{moresco2021} and \citet{veropalumbo2021}
we exploit the same spectroscopic cluster catalog analyzed in this
work, measuring the two-point and three-point correlation functions up
to the BAO scales, and providing new constraints on the geometry of
the universe and on the nonlinear bias of the sample.


\section*{Acknowledgements}

We acknowledge the grants ASI n.I/023/12/0 and ASI n.2018-23-HH.0, and
the use of computational resources from the parallel computing cluster
of the Open Physics Hub
(\href{https://site.unibo.it/openphysicshub/en}{site.unibo.it/openphysicshub/en})
at the Department of Physics and Astronomy, University of
Bologna. L.M. acknowledges support from the grant PRIN-MIUR 2017
WSCC32.

{\em Software}: \texttt{CosmoBolognaLib} \citep{marulli2016};
\texttt{MANGLE} \citep{swanson2008_MANGLE}; \texttt{CAMB}
\citep{lewis2000}; \texttt{CPT Library} \citep{taruya2008, zhao2021};
\texttt{FFTLog} \citep{hamilton2000}; \texttt{Matplotlib}
\citep{hunter2007}.


\bibliography{bib}{}

\begin{thebibliography}{}
\expandafter\ifx\csname natexlab\endcsname\relax\def\natexlab#1{#1}\fi
\providecommand{\url}[1]{\href{#1}{#1}}
\providecommand{\dodoi}[1]{doi:~\href{http://doi.org/#1}{\nolinkurl{#1}}}
\providecommand{\doeprint}[1]{\href{http://ascl.net/#1}{\nolinkurl{http://ascl.net/#1}}}
\providecommand{\doarXiv}[1]{\href{https://arxiv.org/abs/#1}{\nolinkurl{https://arxiv.org/abs/#1}}}

\bibitem[{{Adams} \& {Blake}(2017)}]{adams2017}
{Adams}, C., \& {Blake}, C. 2017, \mnras, 471, 839,
  \dodoi{10.1093/mnras/stx1529}

\bibitem[{{Aihara} {et~al.}(2011){Aihara}, {Allende Prieto}, {An}, {Anderson},
  {Aubourg}, {Balbinot}, {Beers}, {Berlind}, {Bickerton}, {Bizyaev}, {Blanton},
  {Bochanski}, {Bolton}, {Bovy}, {Brandt}, {Brinkmann}, {Brown}, {Brownstein},
  {Busca}, {Campbell}, {Carr}, {Chen}, {Chiappini}, {Comparat}, {Connolly},
  {Cortes}, {Croft}, {Cuesta}, {da Costa}, {Davenport}, {Dawson}, {Dhital},
  {Ealet}, {Ebelke}, {Edmondson}, {Eisenstein}, {Escoffier}, {Esposito},
  {Evans}, {Fan}, {Femen{\'{\i}}a Castell{\'a}}, {Font-Ribera}, {Frinchaboy},
  {Ge}, {Gillespie}, {Gilmore}, {Gonz{\'a}lez Hern{\'a}ndez}, {Gott}, {Gould},
  {Grebel}, {Gunn}, {Hamilton}, {Harding}, {Harris}, {Hawley}, {Hearty}, {Ho},
  {Hogg}, {Holtzman}, {Honscheid}, {Inada}, {Ivans}, {Jiang}, {Johnson},
  {Jordan}, {Jordan}, {Kazin}, {Kirkby}, {Klaene}, {Knapp}, {Kneib},
  {Kochanek}, {Koesterke}, {Kollmeier}, {Kron}, {Lampeitl}, {Lang}, {Le Goff},
  {Lee}, {Lin}, {Long}, {Loomis}, {Lucatello}, {Lundgren}, {Lupton}, {Ma},
  {MacDonald}, {Mahadevan}, {Maia}, {Makler}, {Malanushenko}, {Malanushenko},
  {Mandelbaum}, {Maraston}, {Margala}, {Masters}, {McBride}, {McGehee},
  {McGreer}, {M{\'e}nard}, {Miralda-Escud{\'e}}, {Morrison}, {Mullally},
  {Muna}, {Munn}, {Murayama}, {Myers}, {Naugle}, {Neto}, {Nguyen}, {Nichol},
  {O'Connell}, {Ogando}, {Olmstead}, {Oravetz}, {Padmanabhan},
  {Palanque-Delabrouille}, {Pan}, {Pandey}, {P{\^a}ris}, {Percival},
  {Petitjean}, {Pfaffenberger}, {Pforr}, {Phleps}, {Pichon}, {Pieri}, {Prada},
  {Price-Whelan}, {Raddick}, {Ramos}, {Reyl{\'e}}, {Rich}, {Richards}, {Rix},
  {Robin}, {Rocha-Pinto}, {Rockosi}, {Roe}, {Rollinde}, {Ross}, {Ross},
  {Rossetto}, {S{\'a}nchez}, {Sayres}, {Schlegel}, {Schlesinger}, {Schmidt},
  {Schneider}, {Sheldon}, {Shu}, {Simmerer}, {Simmons}, {Sivarani}, {Snedden},
  {Sobeck}, {Steinmetz}, {Strauss}, {Szalay}, {Tanaka}, {Thakar}, {Thomas},
  {Tinker}, {Tofflemire}, {Tojeiro}, {Tremonti}, {Vandenberg}, {Vargas
  Maga{\~n}a}, {Verde}, {Vogt}, {Wake}, {Wang}, {Weaver}, {Weinberg}, {White},
  {White}, {Yanny}, {Yasuda}, {Yeche}, \& {Zehavi}}]{aihara2011}
{Aihara}, H., {Allende Prieto}, C., {An}, D., {et~al.} 2011, \apjs, 193, 29,
  \dodoi{10.1088/0067-0049/193/2/29}

\bibitem[{{Alam} {et~al.}(2015){Alam}, {Albareti}, {Allende Prieto}, {Anders},
  {Anderson}, {Anderton}, {Andrews}, {Armengaud}, {Aubourg}, {Bailey}, \&
  et~al.}]{alam2015}
{Alam}, S., {Albareti}, F.~D., {Allende Prieto}, C., {et~al.} 2015, \apjs, 219,
  12, \dodoi{10.1088/0067-0049/219/1/12}

\bibitem[{{Alam} {et~al.}(2017){Alam}, {Ata}, {Bailey}, {Beutler}, {Bizyaev},
  {Blazek}, {Bolton}, {Brownstein}, {Burden}, {Chuang}, {Comparat}, {Cuesta},
  {Dawson}, {Eisenstein}, {Escoffier}, {Gil-Mar{\'\i}n}, {Grieb}, {Hand}, {Ho},
  {Kinemuchi}, {Kirkby}, {Kitaura}, {Malanushenko}, {Malanushenko}, {Maraston},
  {McBride}, {Nichol}, {Olmstead}, {Oravetz}, {Padmanabhan},
  {Palanque-Delabrouille}, {Pan}, {Pellejero-Ibanez}, {Percival}, {Petitjean},
  {Prada}, {Price-Whelan}, {Reid}, {Rodr{\'\i}guez-Torres}, {Roe}, {Ross},
  {Ross}, {Rossi}, {Rubi{\~n}o-Mart{\'\i}n}, {Saito}, {Salazar-Albornoz},
  {Samushia}, {S{\'a}nchez}, {Satpathy}, {Schlegel}, {Schneider},
  {Sc{\'o}ccola}, {Seo}, {Sheldon}, {Simmons}, {Slosar}, {Strauss}, {Swanson},
  {Thomas}, {Tinker}, {Tojeiro}, {Maga{\~n}a}, {Vazquez}, {Verde}, {Wake},
  {Wang}, {Weinberg}, {White}, {Wood-Vasey}, {Y{\`e}che}, {Zehavi}, {Zhai}, \&
  {Zhao}}]{alam2017}
{Alam}, S., {Ata}, M., {Bailey}, S., {et~al.} 2017, \mnras, 470, 2617,
  \dodoi{10.1093/mnras/stx721}

\bibitem[{{Alam} {et~al.}(2021){Alam}, {Aubert}, {Avila}, {Balland},
  {Bautista}, {Bershady}, {Bizyaev}, {Blanton}, {Bolton}, {Bovy}, {Brinkmann},
  {Brownstein}, {Burtin}, {Chabanier}, {Chapman}, {Choi}, {Chuang}, {Comparat},
  {Cousinou}, {Cuceu}, {Dawson}, {de la Torre}, {de Mattia}, {Agathe}, {des
  Bourboux}, {Escoffier}, {Etourneau}, {Farr}, {Font-Ribera}, {Frinchaboy},
  {Fromenteau}, {Gil-Mar{\'\i}n}, {Le Goff}, {Gonzalez-Morales},
  {Gonzalez-Perez}, {Grabowski}, {Guy}, {Hawken}, {Hou}, {Kong}, {Parker},
  {Klaene}, {Kneib}, {Lin}, {Long}, {Lyke}, {de la Macorra}, {Martini},
  {Masters}, {Mohammad}, {Moon}, {Mueller}, {Mu{\~n}oz-Guti{\'e}rrez}, {Myers},
  {Nadathur}, {Neveux}, {Newman}, {Noterdaeme}, {Oravetz}, {Oravetz},
  {Palanque-Delabrouille}, {Pan}, {Paviot}, {Percival}, {P{\'e}rez-R{\`a}fols},
  {Petitjean}, {Pieri}, {Prakash}, {Raichoor}, {Ravoux}, {Rezaie}, {Rich},
  {Ross}, {Rossi}, {Ruggeri}, {Ruhlmann-Kleider}, {S{\'a}nchez}, {S{\'a}nchez},
  {S{\'a}nchez-Gallego}, {Sayres}, {Schneider}, {Seo}, {Shafieloo}, {Slosar},
  {Smith}, {Stermer}, {Tamone}, {Tinker}, {Tojeiro}, {Vargas-Maga{\~n}a},
  {Variu}, {Wang}, {Weaver}, {Weijmans}, {Y{\`e}che}, {Zarrouk}, {Zhao},
  {Zhao}, \& {Zheng}}]{eBOSS2021}
{Alam}, S., {Aubert}, M., {Avila}, S., {et~al.} 2021, \prd, 103, 083533,
  \dodoi{10.1103/PhysRevD.103.083533}

\bibitem[{{Alcock} \& {Paczynski}(1979)}]{alcock1979}
{Alcock}, C., \& {Paczynski}, B. 1979, \nat, 281, 358, \dodoi{10.1038/281358a0}

\bibitem[{{Amendola}(2000)}]{amendola2000}
{Amendola}, L. 2000, \prd, 62, 043511, \dodoi{10.1103/PhysRevD.62.043511}

\bibitem[{{Amendola} {et~al.}(2018){Amendola}, {Appleby}, {Avgoustidis},
  {Bacon}, {Baker}, {Baldi}, {Bartolo}, {Blanchard}, {Bonvin}, {Borgani},
  {Branchini}, {Burrage}, {Camera}, {Carbone}, {Casarini}, {Cropper}, {de
  Rham}, {Dietrich}, {Di Porto}, {Durrer}, {Ealet}, {Ferreira}, {Finelli},
  {Garc{\'{\i}}a-Bellido}, {Giannantonio}, {Guzzo}, {Heavens}, {Heisenberg},
  {Heymans}, {Hoekstra}, {Hollenstein}, {Holmes}, {Hwang}, {Jahnke},
  {Kitching}, {Koivisto}, {Kunz}, {La Vacca}, {Linder}, {March}, {Marra},
  {Martins}, {Majerotto}, {Markovic}, {Marsh}, {Marulli}, {Massey}, {Mellier},
  {Montanari}, {Mota}, {Nunes}, {Percival}, {Pettorino}, {Porciani},
  {Quercellini}, {Read}, {Rinaldi}, {Sapone}, {Sawicki}, {Scaramella},
  {Skordis}, {Simpson}, {Taylor}, {Thomas}, {Trotta}, {Verde}, {Vernizzi},
  {Vollmer}, {Wang}, {Weller}, \& {Zlosnik}}]{amendola2018}
{Amendola}, L., {Appleby}, S., {Avgoustidis}, A., {et~al.} 2018, Living Reviews
  in Relativity, 21, 2, \dodoi{10.1007/s41114-017-0010-3}

\bibitem[{{Angulo} {et~al.}(2005){Angulo}, {Baugh}, {Frenk}, {Bower},
  {Jenkins}, \& {Morris}}]{angulo2005}
{Angulo}, R.~E., {Baugh}, C.~M., {Frenk}, C.~S., {et~al.} 2005, \mnras, 362,
  L25, \dodoi{10.1111/j.1745-3933.2005.00067.x}

\bibitem[{{Aubert} {et~al.}(2020){Aubert}, {Cousinou}, {Escoffier}, {Hawken},
  {Nadathur}, {Alam}, {Bautista}, {Burtin}, {de Mattia}, {Gil-Mar{\'\i}n},
  {Hou}, {Jullo}, {Neveux}, {Rossi}, {Smith}, {Tamone}, \& {Vargas
  Maga{\~n}a}}]{aubert2020}
{Aubert}, M., {Cousinou}, M.-C., {Escoffier}, S., {et~al.} 2020, arXiv
  e-prints, arXiv:2007.09013.
\newblock \doarXiv{2007.09013}

\bibitem[{{Balaguera-Antol{\'{\i}}nez}
  {et~al.}(2011){Balaguera-Antol{\'{\i}}nez}, {S{\'a}nchez}, {B{\"o}hringer},
  {Collins}, {Guzzo}, \& {Phleps}}]{balaguera2011}
{Balaguera-Antol{\'{\i}}nez}, A., {S{\'a}nchez}, A.~G., {B{\"o}hringer}, H.,
  {et~al.} 2011, \mnras, 413, 386, \dodoi{10.1111/j.1365-2966.2010.18143.x}

\bibitem[{{Bautista} {et~al.}(2021){Bautista}, {Paviot}, {Vargas Maga{\~n}a},
  {de la Torre}, {Fromenteau}, {Gil-Mar{\'\i}n}, {Ross}, {Burtin}, {Dawson},
  {Hou}, {Kneib}, {de Mattia}, {Percival}, {Rossi}, {Tojeiro}, {Zhao}, {Zhao},
  {Alam}, {Brownstein}, {Chapman}, {Choi}, {Chuang}, {Escoffier}, {de la
  Macorra}, {du Mas des Bourboux}, {Mohammad}, {Moon}, {M{\"u}ller},
  {Nadathur}, {Newman}, {Schneider}, {Seo}, \& {Wang}}]{bautista2021}
{Bautista}, J.~E., {Paviot}, R., {Vargas Maga{\~n}a}, M., {et~al.} 2021,
  \mnras, 500, 736, \dodoi{10.1093/mnras/staa2800}

\bibitem[{{Bel} {et~al.}(2019){Bel}, {Pezzotta}, {Carbone}, {Sefusatti}, \&
  {Guzzo}}]{bel2019}
{Bel}, J., {Pezzotta}, A., {Carbone}, C., {Sefusatti}, E., \& {Guzzo}, L. 2019,
  \aap, 622, A109, \dodoi{10.1051/0004-6361/201834513}

\bibitem[{{Berlind} {et~al.}(2006){Berlind}, {Frieman}, {Weinberg}, {Blanton},
  {Warren}, {Abazajian}, {Scranton}, {Hogg}, {Scoccimarro}, {Bahcall},
  {Brinkmann}, {Gott}, {Kleinman}, {Krzesinski}, {Lee}, {Miller}, {Nitta},
  {Schneider}, {Tucker}, {Zehavi}, \& {SDSS Collaboration}}]{berlind2006}
{Berlind}, A.~A., {Frieman}, J., {Weinberg}, D.~H., {et~al.} 2006, \apjs, 167,
  1, \dodoi{10.1086/508170}

\bibitem[{{Beutler} {et~al.}(2012){Beutler}, {Blake}, {Colless}, {Jones},
  {Staveley-Smith}, {Poole}, {Campbell}, {Parker}, {Saunders}, \&
  {Watson}}]{beutler2012}
{Beutler}, F., {Blake}, C., {Colless}, M., {et~al.} 2012, \mnras, 423, 3430,
  \dodoi{10.1111/j.1365-2966.2012.21136.x}

\bibitem[{{Beutler} {et~al.}(2014){Beutler}, {Saito}, {Seo}, {Brinkmann},
  {Dawson}, {Eisenstein}, {Font-Ribera}, {Ho}, {McBride}, \&
  {Montesano}}]{beutler2014}
{Beutler}, F., {Saito}, S., {Seo}, H.-J., {et~al.} 2014, \mnras, 443, 1065,
  \dodoi{10.1093/mnras/stu1051}

\bibitem[{{Beutler} {et~al.}(2017){Beutler}, {Seo}, {Saito}, {Chuang},
  {Cuesta}, {Eisenstein}, {Gil-Mar{\'\i}n}, {Grieb}, {Hand}, {Kitaura}, {Modi},
  {Nichol}, {Olmstead}, {Percival}, {Prada}, {S{\'a}nchez}, {Rodriguez-Torres},
  {Ross}, {Ross}, {Schneider}, {Tinker}, {Tojeiro}, \&
  {Vargas-Maga{\~n}a}}]{beutler2017}
{Beutler}, F., {Seo}, H.-J., {Saito}, S., {et~al.} 2017, \mnras, 466, 2242,
  \dodoi{10.1093/mnras/stw3298}

\bibitem[{{Blake} {et~al.}(2012){Blake}, {Brough}, {Colless}, {Contreras},
  {Couch}, {Croom}, {Croton}, {Davis}, {Drinkwater}, {Forster}, {Gilbank},
  {Gladders}, {Glazebrook}, {Jelliffe}, {Jurek}, {Li}, {Madore}, {Martin},
  {Pimbblet}, {Poole}, {Pracy}, {Sharp}, {Wisnioski}, {Woods}, {Wyder}, \&
  {Yee}}]{blake2012}
{Blake}, C., {Brough}, S., {Colless}, M., {et~al.} 2012, \mnras, 425, 405,
  \dodoi{10.1111/j.1365-2966.2012.21473.x}

\bibitem[{{Blake} {et~al.}(2013){Blake}, {Baldry}, {Bland-Hawthorn},
  {Christodoulou}, {Colless}, {Conselice}, {Driver}, {Hopkins}, {Liske},
  {Loveday}, {Norberg}, {Peacock}, {Poole}, \& {Robotham}}]{blake2013}
{Blake}, C., {Baldry}, I.~K., {Bland-Hawthorn}, J., {et~al.} 2013, \mnras, 436,
  3089, \dodoi{10.1093/mnras/stt1791}

\bibitem[{{Blanton} {et~al.}(2003){Blanton}, {Hogg}, {Bahcall}, {Brinkmann},
  {Britton}, {Connolly}, {Csabai}, {Fukugita}, {Loveday}, {Meiksin}, {Munn},
  {Nichol}, {Okamura}, {Quinn}, {Schneider}, {Shimasaku}, {Strauss}, {Tegmark},
  {Vogeley}, \& {Weinberg}}]{blanton2003}
{Blanton}, M.~R., {Hogg}, D.~W., {Bahcall}, N.~A., {et~al.} 2003, \apj, 592,
  819, \dodoi{10.1086/375776}

\bibitem[{{Chan} {et~al.}(2012){Chan}, {Scoccimarro}, \& {Sheth}}]{chan2012}
{Chan}, K.~C., {Scoccimarro}, R., \& {Sheth}, R.~K. 2012, \prd, 85, 083509,
  \dodoi{10.1103/PhysRevD.85.083509}

\bibitem[{{Chuang} \& {Wang}(2013)}]{chuang2013}
{Chuang}, C.-H., \& {Wang}, Y. 2013, \mnras, 435, 255,
  \dodoi{10.1093/mnras/stt1290}

\bibitem[{{Chuang} {et~al.}(2013){Chuang}, {Prada}, {Cuesta}, {Eisenstein},
  {Kazin}, {Padmanabhan}, {S{\'a}nchez}, {Xu}, {Beutler}, {Manera}, {Schlegel},
  {Schneider}, {Weinberg}, {Brinkmann}, {Brownstein}, \&
  {Thomas}}]{chuang2013b}
{Chuang}, C.-H., {Prada}, F., {Cuesta}, A.~J., {et~al.} 2013, \mnras, 433,
  3559, \dodoi{10.1093/mnras/stt988}

\bibitem[{{Chuang} {et~al.}(2016){Chuang}, {Prada}, {Pellejero-Ibanez},
  {Beutler}, {Cuesta}, {Eisenstein}, {Escoffier}, {Ho}, {Kitaura}, \&
  {Kneib}}]{chuang2016}
{Chuang}, C.-H., {Prada}, F., {Pellejero-Ibanez}, M., {et~al.} 2016, \mnras,
  461, 3781, \dodoi{10.1093/mnras/stw1535}

\bibitem[{{Costanzi} {et~al.}(2019){Costanzi}, {Rozo}, {Simet}, {Zhang},
  {Evrard}, {Mantz}, {Rykoff}, {Jeltema}, {Gruen}, {Allen}, {McClintock},
  {Romer}, {von der Linden}, {Farahi}, {DeRose}, {Varga}, {Weller}, {Giles},
  {Hollowood}, {Bhargava}, {Bermeo-Hernandez}, {Chen}, {Abbott}, {Abdalla},
  {Avila}, {Bechtol}, {Brooks}, {Buckley-Geer}, {Burke}, {Rosell}, {Kind},
  {Carretero}, {Crocce}, {Cunha}, {da Costa}, {Davis}, {De Vicente}, {Diehl},
  {Dietrich}, {Doel}, {Eifler}, {Estrada}, {Flaugher}, {Fosalba}, {Frieman},
  {Garc{\'\i}a-Bellido}, {Gaztanaga}, {Gerdes}, {Giannantonio}, {Gruendl},
  {Gschwend}, {Gutierrez}, {Hartley}, {Honscheid}, {Hoyle}, {James}, {Krause},
  {Kuehn}, {Kuropatkin}, {Lima}, {Lin}, {Maia}, {March}, {Marshall}, {Martini},
  {Menanteau}, {Miller}, {Miquel}, {Mohr}, {Ogando}, {Plazas}, {Roodman},
  {Sanchez}, {Scarpine}, {Schindler}, {Schubnell}, {Serrano}, {Sevilla-Noarbe},
  {Sheldon}, {Smith}, {Soares-Santos}, {Sobreira}, {Suchyta}, {Swanson},
  {Tarle}, {Thomas}, \& {Wechsler}}]{costanzi2019}
{Costanzi}, M., {Rozo}, E., {Simet}, M., {et~al.} 2019, \mnras, 488, 4779,
  \dodoi{10.1093/mnras/stz1949}

\bibitem[{{Covone} {et~al.}(2014){Covone}, {Sereno}, {Kilbinger}, \&
  {Cardone}}]{covone2014}
{Covone}, G., {Sereno}, M., {Kilbinger}, M., \& {Cardone}, V.~F. 2014, \apjl,
  784, L25, \dodoi{10.1088/2041-8205/784/2/L25}

\bibitem[{{Davis} {et~al.}(2011){Davis}, {Nusser}, {Masters}, {Springob},
  {Huchra}, \& {Lemson}}]{davis2011}
{Davis}, M., {Nusser}, A., {Masters}, K.~L., {et~al.} 2011, \mnras, 413, 2906,
  \dodoi{10.1111/j.1365-2966.2011.18362.x}

\bibitem[{{Davis} \& {Peebles}(1983)}]{davis1983}
{Davis}, M., \& {Peebles}, P.~J.~E. 1983, \apj, 267, 465,
  \dodoi{10.1086/160884}

\bibitem[{{Dawson} {et~al.}(2013){Dawson}, {Schlegel}, {Ahn}, {Anderson},
  {Aubourg}, {Bailey}, {Barkhouser}, {Bautista}, {Beifiori}, {Berlind},
  {Bhardwaj}, {Bizyaev}, {Blake}, {Blanton}, {Blomqvist}, {Bolton}, {Borde},
  {Bovy}, {Brandt}, {Brewington}, {Brinkmann}, {Brown}, {Brownstein}, {Bundy},
  {Busca}, {Carithers}, {Carnero}, {Carr}, {Chen}, {Comparat}, {Connolly},
  {Cope}, {Croft}, {Cuesta}, {da Costa}, {Davenport}, {Delubac}, {de Putter},
  {Dhital}, {Ealet}, {Ebelke}, {Eisenstein}, {Escoffier}, {Fan}, {Filiz Ak},
  {Finley}, {Font-Ribera}, {G{\'e}nova-Santos}, {Gunn}, {Guo}, {Haggard},
  {Hall}, {Hamilton}, {Harris}, {Harris}, {Ho}, {Hogg}, {Holder}, {Honscheid},
  {Huehnerhoff}, {Jordan}, {Jordan}, {Kauffmann}, {Kazin}, {Kirkby}, {Klaene},
  {Kneib}, {Le Goff}, {Lee}, {Long}, {Loomis}, {Lundgren}, {Lupton}, {Maia},
  {Makler}, {Malanushenko}, {Malanushenko}, {Mandelbaum}, {Manera}, {Maraston},
  {Margala}, {Masters}, {McBride}, {McDonald}, {McGreer}, {McMahon}, {Mena},
  {Miralda-Escud{\'e}}, {Montero-Dorta}, {Montesano}, {Muna}, {Myers},
  {Naugle}, {Nichol}, {Noterdaeme}, {Nuza}, {Olmstead}, {Oravetz}, {Oravetz},
  {Owen}, {Padmanabhan}, {Palanque-Delabrouille}, {Pan}, {Parejko},
  {P{\^a}ris}, {Percival}, {P{\'e}rez-Fournon}, {P{\'e}rez-R{\`a}fols},
  {Petitjean}, {Pfaffenberger}, {Pforr}, {Pieri}, {Prada}, {Price-Whelan},
  {Raddick}, {Rebolo}, {Rich}, {Richards}, {Rockosi}, {Roe}, {Ross}, {Ross},
  {Rossi}, {Rubi{\~n}o-Martin}, {Samushia}, {S{\'a}nchez}, {Sayres}, {Schmidt},
  {Schneider}, {Sc{\'o}ccola}, {Seo}, {Shelden}, {Sheldon}, {Shen}, {Shu},
  {Slosar}, {Smee}, {Snedden}, {Stauffer}, {Steele}, {Strauss}, {Streblyanska},
  {Suzuki}, {Swanson}, {Tal}, {Tanaka}, {Thomas}, {Tinker}, {Tojeiro},
  {Tremonti}, {Vargas Maga{\~n}a}, {Verde}, {Viel}, {Wake}, {Watson}, {Weaver},
  {Weinberg}, {Weiner}, {West}, {White}, {Wood-Vasey}, {Yeche}, {Zehavi},
  {Zhao}, \& {Zheng}}]{dawson2013}
{Dawson}, K.~S., {Schlegel}, D.~J., {Ahn}, C.~P., {et~al.} 2013, \aj, 145, 10,
  \dodoi{10.1088/0004-6256/145/1/10}

\bibitem[{{De Felice} \& {Tsujikawa}(2010)}]{defelice2010}
{De Felice}, A., \& {Tsujikawa}, S. 2010, Living Reviews in Relativity, 13, 3,
  \dodoi{10.12942/lrr-2010-3}

\bibitem[{{de la Torre} \& {Guzzo}(2012)}]{delatorre2012}
{de la Torre}, S., \& {Guzzo}, L. 2012, \mnras, 427, 327,
  \dodoi{10.1111/j.1365-2966.2012.21824.x}

\bibitem[{{de la Torre} {et~al.}(2013){de la Torre}, {Guzzo}, {Peacock},
  {Branchini}, {Iovino}, {Granett}, {Abbas}, {Adami}, {Arnouts}, {Bel},
  {Bolzonella}, {Bottini}, {Cappi}, {Coupon}, {Cucciati}, {Davidzon}, {De
  Lucia}, {Fritz}, {Franzetti}, {Fumana}, {Garilli}, {Ilbert}, {Krywult}, {Le
  Brun}, {Le F{\`e}vre}, {Maccagni}, {Ma{\l}ek}, {Marulli}, {McCracken},
  {Moscardini}, {Paioro}, {Percival}, {Polletta}, {Pollo}, {Schlagenhaufer},
  {Scodeggio}, {Tasca}, {Tojeiro}, {Vergani}, {Zanichelli}, {Burden}, {Di
  Porto}, {Marchetti}, {Marinoni}, {Mellier}, {Monaco}, {Nichol}, {Phleps},
  {Wolk}, \& {Zamorani}}]{delatorre2013b}
{de la Torre}, S., {Guzzo}, L., {Peacock}, J.~A., {et~al.} 2013, \aap, 557,
  A54, \dodoi{10.1051/0004-6361/201321463}

\bibitem[{{de la Torre} {et~al.}(2017){de la Torre}, {Jullo}, {Giocoli},
  {Pezzotta}, {Bel}, {Granett}, {Guzzo}, {Garilli}, {Scodeggio}, {Bolzonella},
  {Abbas}, {Adami}, {Bottini}, {Cappi}, {Cucciati}, {Davidzon}, {Franzetti},
  {Fritz}, {Iovino}, {Krywult}, {Le Brun}, {Le F{\`e}vre}, {Maccagni},
  {Ma{\l}ek}, {Marulli}, {Polletta}, {Pollo}, {Tasca}, {Tojeiro}, {Vergani},
  {Zanichelli}, {Arnouts}, {Branchini}, {Coupon}, {De Lucia}, {Ilbert},
  {Moutard}, {Moscardini}, {Peacock}, {Metcalf}, {Prada}, \&
  {Yepes}}]{delatorre2017}
{de la Torre}, S., {Jullo}, E., {Giocoli}, C., {et~al.} 2017, \aap, 608, A44,
  \dodoi{10.1051/0004-6361/201630276}

\bibitem[{{de Mattia} {et~al.}(2021){de Mattia}, {Ruhlmann-Kleider},
  {Raichoor}, {Ross}, {Tamone}, {Zhao}, {Alam}, {Avila}, {Burtin}, {Bautista},
  {Beutler}, {Brinkmann}, {Brownstein}, {Chapman}, {Chuang}, {Comparat}, {du
  Mas des Bourboux}, {Dawson}, {de la Macorra}, {Gil-Mar{\'\i}n},
  {Gonzalez-Perez}, {Gorgoni}, {Hou}, {Kong}, {Lin}, {Nadathur}, {Newman},
  {Mueller}, {Percival}, {Rezaie}, {Rossi}, {Schneider}, {Tiwari}, {Vivek},
  {Wang}, \& {Zhao}}]{demattia2021}
{de Mattia}, A., {Ruhlmann-Kleider}, V., {Raichoor}, A., {et~al.} 2021, \mnras,
  501, 5616, \dodoi{10.1093/mnras/staa3891}

\bibitem[{{Desjacques} {et~al.}(2018){Desjacques}, {Jeong}, \&
  {Schmidt}}]{desjacques2018}
{Desjacques}, V., {Jeong}, D., \& {Schmidt}, F. 2018, \physrep, 733, 1,
  \dodoi{10.1016/j.physrep.2017.12.002}

\bibitem[{{di Porto} \& {Amendola}(2008)}]{diporto2008}
{di Porto}, C., \& {Amendola}, L. 2008, \prd, 77, 083508,
  \dodoi{10.1103/PhysRevD.77.083508}

\bibitem[{{Di Porto} {et~al.}(2012){Di Porto}, {Amendola}, \&
  {Branchini}}]{diporto2012}
{Di Porto}, C., {Amendola}, L., \& {Branchini}, E. 2012, \mnras, 419, 985,
  \dodoi{10.1111/j.1365-2966.2011.19755.x}

\bibitem[{{Dvali} {et~al.}(2000){Dvali}, {Gabadadze}, \& {Porrati}}]{Dvali2000}
{Dvali}, G., {Gabadadze}, G., \& {Porrati}, M. 2000, Physics Letters B, 485,
  208, \dodoi{10.1016/S0370-2693(00)00669-9}

\bibitem[{{Emami} {et~al.}(2017){Emami}, {Broadhurst}, {Jimeno}, {Smoot},
  {Angulo}, {Lim}, {Chung Chu}, \& {Lazkoz}}]{emami2017}
{Emami}, R., {Broadhurst}, T., {Jimeno}, P., {et~al.} 2017, ArXiv e-prints:
  1711.05210.
\newblock \doarXiv{1711.05210}

\bibitem[{{Estrada} {et~al.}(2009){Estrada}, {Sefusatti}, \&
  {Frieman}}]{estrada2009}
{Estrada}, J., {Sefusatti}, E., \& {Frieman}, J.~A. 2009, \apj, 692, 265,
  \dodoi{10.1088/0004-637X/692/1/265}

\bibitem[{{Feix} {et~al.}(2015){Feix}, {Nusser}, \& {Branchini}}]{feix2015}
{Feix}, M., {Nusser}, A., \& {Branchini}, E. 2015, \prl, 115, 011301,
  \dodoi{10.1103/PhysRevLett.115.011301}

\bibitem[{{Fisher} {et~al.}(1994){Fisher}, {Scharf}, \& {Lahav}}]{fisher1994}
{Fisher}, K.~B., {Scharf}, C.~A., \& {Lahav}, O. 1994, \mnras, 266, 219

\bibitem[{{Fu} {et~al.}(2009){Fu}, {Wu}, \& {Yu}}]{fu2009}
{Fu}, X., {Wu}, P., \& {Yu}, H. 2009, Physics Letters B, 677, 12,
  \dodoi{10.1016/j.physletb.2009.05.007}

\bibitem[{{Gannouji} {et~al.}(2009){Gannouji}, {Moraes}, \&
  {Polarski}}]{gannouji2009}
{Gannouji}, R., {Moraes}, B., \& {Polarski}, D. 2009, \jcap, 2009, 034,
  \dodoi{10.1088/1475-7516/2009/02/034}

\bibitem[{{Garc{\'{\i}}a-Farieta} {et~al.}(2020){Garc{\'{\i}}a-Farieta},
  {Marulli}, {Moscardini}, {Veropalumbo}, \& {Casas-Mirand
  a}}]{garcia-farieta2020}
{Garc{\'{\i}}a-Farieta}, J.~E., {Marulli}, F., {Moscardini}, L., {Veropalumbo},
  A., \& {Casas-Mirand a}, R.~A. 2020, \mnras, 494, 1658,
  \dodoi{10.1093/mnras/staa791}

\bibitem[{{Garc{\'{\i}}a-Farieta} {et~al.}(2019){Garc{\'{\i}}a-Farieta},
  {Marulli}, {Veropalumbo}, {Moscardini}, {Casas-Miranda}, {Giocoli}, \&
  {Baldi}}]{garcia-farieta2019}
{Garc{\'{\i}}a-Farieta}, J.~E., {Marulli}, F., {Veropalumbo}, A., {et~al.}
  2019, \mnras, 488, 1987, \dodoi{10.1093/mnras/stz1850}

\bibitem[{{Garc{\'{\i}}a-Farieta} {et~al.}(in prep.)}]{garcia-farieta2021}
{Garc{\'{\i}}a-Farieta}, J.~E., {et~al.} in prep.

\bibitem[{{Gil-Mar{\'\i}n} {et~al.}(2014){Gil-Mar{\'\i}n}, {Wagner},
  {Nore{\~n}a}, {Verde}, \& {Percival}}]{gilmarin2014}
{Gil-Mar{\'\i}n}, H., {Wagner}, C., {Nore{\~n}a}, J., {Verde}, L., \&
  {Percival}, W. 2014, \jcap, 2014, 029, \dodoi{10.1088/1475-7516/2014/12/029}

\bibitem[{{Gil-Mar{\'\i}n} {et~al.}(2012){Gil-Mar{\'\i}n}, {Wagner}, {Verde},
  {Porciani}, \& {Jimenez}}]{gilmarin2012}
{Gil-Mar{\'\i}n}, H., {Wagner}, C., {Verde}, L., {Porciani}, C., \& {Jimenez},
  R. 2012, Journal of Cosmology and Astro-Particle Physics, 2012, 029,
  \dodoi{10.1088/1475-7516/2012/11/029}

\bibitem[{{Gil-Mar{\'\i}n} {et~al.}(2020){Gil-Mar{\'\i}n}, {Bautista},
  {Paviot}, {Vargas-Maga{\~n}a}, {de la Torre}, {Fromenteau}, {Alam},
  {{\'A}vila}, {Burtin}, {Chuang}, {Dawson}, {Hou}, {de Mattia}, {Mohammad},
  {M{\"u}ller}, {Nadathur}, {Neveux}, {Percival}, {Raichoor}, {Rezaie}, {Ross},
  {Rossi}, {Ruhlmann-Kleider}, {Smith}, {Tamone}, {Tinker}, {Tojeiro}, {Wang},
  {Zhao}, {Zhao}, {Brinkmann}, {Brownstein}, {Choi}, {Escoffier}, {de la
  Macorra}, {Moon}, {Newman}, {Schneider}, {Seo}, \& {Vivek}}]{gilmarin2020}
{Gil-Mar{\'\i}n}, H., {Bautista}, J.~E., {Paviot}, R., {et~al.} 2020, \mnras,
  498, 2492, \dodoi{10.1093/mnras/staa2455}

\bibitem[{{Grieb} {et~al.}(2016){Grieb}, {S{\'a}nchez}, {Salazar-Albornoz}, \&
  {Dalla Vecchia}}]{grieb2016}
{Grieb}, J.~N., {S{\'a}nchez}, A.~G., {Salazar-Albornoz}, S., \& {Dalla
  Vecchia}, C. 2016, \mnras, 457, 1577, \dodoi{10.1093/mnras/stw065}

\bibitem[{{Guzzo} {et~al.}(2008){Guzzo}, {Pierleoni}, {Meneux}, {Branchini},
  {Le F{\`e}vre}, {Marinoni}, {Garilli}, {Blaizot}, {De Lucia}, \&
  {Pollo}}]{guzzo2008}
{Guzzo}, L., {Pierleoni}, M., {Meneux}, B., {et~al.} 2008, \nat, 451, 541,
  \dodoi{10.1038/nature06555}

\bibitem[{{Hamaus} {et~al.}(2020){Hamaus}, {Pisani}, {Choi}, {Lavaux},
  {Wandelt}, \& {Weller}}]{hamaus2020}
{Hamaus}, N., {Pisani}, A., {Choi}, J.-A., {et~al.} 2020, \jcap, 2020, 023,
  \dodoi{10.1088/1475-7516/2020/12/023}

\bibitem[{{Hamaus} {et~al.}(2016){Hamaus}, {Pisani}, {Sutter}, {Lavaux},
  {Escoffier}, {Wand elt}, \& {Weller}}]{hamaus2016}
{Hamaus}, N., {Pisani}, A., {Sutter}, P.~M., {et~al.} 2016, \prl, 117, 091302,
  \dodoi{10.1103/PhysRevLett.117.091302}

\bibitem[{{Hamilton}(1998)}]{hamilton1998}
{Hamilton}, A.~J.~S. 1998, in Astrophysics and Space Science Library, Vol. 231,
  The Evolving Universe, ed. {D.~Hamilton}, 185

\bibitem[{{Hamilton}(2000)}]{hamilton2000}
{Hamilton}, A.~J.~S. 2000, \mnras, 312, 257,
  \dodoi{10.1046/j.1365-8711.2000.03071.x}

\bibitem[{{Hartlap} {et~al.}(2007){Hartlap}, {Simon}, \&
  {Schneider}}]{hartlap2007}
{Hartlap}, J., {Simon}, P., \& {Schneider}, P. 2007, \aap, 464, 399,
  \dodoi{10.1051/0004-6361:20066170}

\bibitem[{{Hawken} {et~al.}(2020){Hawken}, {Aubert}, {Pisani}, {Cousinou},
  {Escoffier}, {Nadathur}, {Rossi}, \& {Schneider}}]{hawken2020}
{Hawken}, A.~J., {Aubert}, M., {Pisani}, A., {et~al.} 2020, \jcap, 2020, 012,
  \dodoi{10.1088/1475-7516/2020/06/012}

\bibitem[{{Hawken} {et~al.}(2017){Hawken}, {Granett}, {Iovino}, {Guzzo},
  {Peacock}, {de la Torre}, {Garilli}, {Bolzonella}, {Scodeggio}, {Abbas},
  {Adami}, {Bottini}, {Cappi}, {Cucciati}, {Davidzon}, {Fritz}, {Franzetti},
  {Krywult}, {Le Brun}, {Le F{\`e}vre}, {Maccagni}, {Ma{\l}ek}, {Marulli},
  {Polletta}, {Pollo}, {Tasca}, {Tojeiro}, {Vergani}, {Zanichelli}, {Arnouts},
  {Bel}, {Branchini}, {De Lucia}, {Ilbert}, {Moscardini}, \&
  {Percival}}]{hawken2017}
{Hawken}, A.~J., {Granett}, B.~R., {Iovino}, A., {et~al.} 2017, \aap, 607, A54,
  \dodoi{10.1051/0004-6361/201629678}

\bibitem[{{Hawkins} {et~al.}(2003){Hawkins}, {Maddox}, {Cole}, {Lahav},
  {Madgwick}, {Norberg}, {Peacock}, {Baldry}, {Baugh}, {Bland-Hawthorn},
  {Bridges}, {Cannon}, {Colless}, {Collins}, {Couch}, {Dalton}, {De Propris},
  {Driver}, {Efstathiou}, {Ellis}, {Frenk}, {Glazebrook}, {Jackson}, {Jones},
  {Lewis}, {Lumsden}, {Percival}, {Peterson}, {Sutherland}, \&
  {Taylor}}]{hawkins2003}
{Hawkins}, E., {Maddox}, S., {Cole}, S., {et~al.} 2003, \mnras, 346, 78,
  \dodoi{10.1046/j.1365-2966.2003.07063.x}

\bibitem[{{Hong} {et~al.}(2016){Hong}, {Han}, \& {Wen}}]{hong2016}
{Hong}, T., {Han}, J.~L., \& {Wen}, Z.~L. 2016, \apj, 826, 154,
  \dodoi{10.3847/0004-637X/826/2/154}

\bibitem[{{Hong} {et~al.}(2012){Hong}, {Han}, {Wen}, {Sun}, \&
  {Zhan}}]{hong2012}
{Hong}, T., {Han}, J.~L., {Wen}, Z.~L., {Sun}, L., \& {Zhan}, H. 2012, \apj,
  749, 81, \dodoi{10.1088/0004-637X/749/1/81}

\bibitem[{{Hou} {et~al.}(2021){Hou}, {S{\'a}nchez}, {Ross}, {Smith}, {Neveux},
  {Bautista}, {Burtin}, {Zhao}, {Scoccimarro}, {Dawson}, {de Mattia}, {de la
  Macorra}, {du Mas des Bourboux}, {Eisenstein}, {Gil-Mar{\'\i}n}, {Lyke},
  {Mohammad}, {Mueller}, {Percival}, {Rossi}, {Vargas Maga{\~n}a}, {Zarrouk},
  {Zhao}, {Brinkmann}, {Brownstein}, {Chuang}, {Myers}, {Newman}, {Schneider},
  \& {Vivek}}]{hou2021}
{Hou}, J., {S{\'a}nchez}, A.~G., {Ross}, A.~J., {et~al.} 2021, \mnras, 500,
  1201, \dodoi{10.1093/mnras/staa3234}

\bibitem[{{Howlett} {et~al.}(2015){Howlett}, {Ross}, {Samushia}, {Percival}, \&
  {Manera}}]{howlett2015}
{Howlett}, C., {Ross}, A.~J., {Samushia}, L., {Percival}, W.~J., \& {Manera},
  M. 2015, \mnras, 449, 848, \dodoi{10.1093/mnras/stu2693}

\bibitem[{{Hu} \& {Sawicki}(2007)}]{hu2007}
{Hu}, W., \& {Sawicki}, I. 2007, \prd, 76, 064004,
  \dodoi{10.1103/PhysRevD.76.064004}

\bibitem[{{Huchra} \& {Geller}(1982)}]{huchra1982}
{Huchra}, J.~P., \& {Geller}, M.~J. 1982, \apj, 257, 423,
  \dodoi{10.1086/160000}

\bibitem[{{Hudson} \& {Turnbull}(2012)}]{hudson2012}
{Hudson}, M.~J., \& {Turnbull}, S.~J. 2012, \apjl, 751, L30,
  \dodoi{10.1088/2041-8205/751/2/L30}

\bibitem[{{Hunter}(2007)}]{hunter2007}
{Hunter}, J.~D. 2007, Computing in Science and Engineering, 9, 90,
  \dodoi{10.1109/MCSE.2007.55}

\bibitem[{{Huterer} {et~al.}(2017){Huterer}, {Shafer}, {Scolnic}, \&
  {Schmidt}}]{huterer2017}
{Huterer}, D., {Shafer}, D.~L., {Scolnic}, D.~M., \& {Schmidt}, F. 2017, \jcap,
  2017, 015, \dodoi{10.1088/1475-7516/2017/05/015}

\bibitem[{{H{\"u}tsi}(2010)}]{hutsi2010}
{H{\"u}tsi}, G. 2010, \mnras, 401, 2477,
  \dodoi{10.1111/j.1365-2966.2009.15824.x}

\bibitem[{{Icaza-Lizaola} {et~al.}(2020){Icaza-Lizaola}, {Vargas-Maga{\~n}a},
  {Fromenteau}, {Alam}, {Camacho}, {Gil-Marin}, {Paviot}, {Ross}, {Schneider},
  {Tinker}, {Wang}, {Zhao}, {Prakash}, {Rossi}, {Zao}, {Cruz-Gonzalez}, \& {de
  la Macorra}}]{icaza-lizaola2020}
{Icaza-Lizaola}, M., {Vargas-Maga{\~n}a}, M., {Fromenteau}, S., {et~al.} 2020,
  \mnras, 492, 4189, \dodoi{10.1093/mnras/stz3602}

\bibitem[{{Kaiser}(1987)}]{kaiser1987}
{Kaiser}, N. 1987, \mnras, 227, 1

\bibitem[{{Kazin} {et~al.}(2012){Kazin}, {S{\'a}nchez}, \&
  {Blanton}}]{kazin2012}
{Kazin}, E.~A., {S{\'a}nchez}, A.~G., \& {Blanton}, M.~R. 2012, \mnras, 419,
  3223, \dodoi{10.1111/j.1365-2966.2011.19962.x}

\bibitem[{{Kazin} {et~al.}(2010){Kazin}, {Blanton}, {Scoccimarro}, {McBride},
  {Berlind}, {Bahcall}, {Brinkmann}, {Czarapata}, {Frieman}, {Kent},
  {Schneider}, \& {Szalay}}]{kazin2010}
{Kazin}, E.~A., {Blanton}, M.~R., {Scoccimarro}, R., {et~al.} 2010, \apj, 710,
  1444, \dodoi{10.1088/0004-637X/710/2/1444}

\bibitem[{{Keih{\"a}nen} {et~al.}(2019){Keih{\"a}nen}, {Kurki-Suonio},
  {Lindholm}, {Viitanen}, {Suur-Uski}, {Allevato}, {Branchini}, {Marulli},
  {Norberg}, {Tavagnacco}, {de la Torre}, {Valiviita}, {Viel}, {Bel},
  {Frailis}, \& {S{\'a}nchez}}]{keihanen2019}
{Keih{\"a}nen}, E., {Kurki-Suonio}, H., {Lindholm}, V., {et~al.} 2019, \aap,
  631, A73, \dodoi{10.1051/0004-6361/201935828}

\bibitem[{{Landy} \& {Szalay}(1993)}]{landy1993}
{Landy}, S.~D., \& {Szalay}, A.~S. 1993, \apj, 412, 64, \dodoi{10.1086/172900}

\bibitem[{{Laureijs} {et~al.}(2011){Laureijs}, {Amiaux}, {Arduini},
  {Augu{\`e}res}, {Brinchmann}, {Cole}, {Cropper}, {Dabin}, {Duvet}, {Ealet},
  {Garilli}, {Gondoin}, {Guzzo}, {Hoar}, {Hoekstra}, {Holmes}, {Kitching},
  {Maciaszek}, {Mellier}, {Pasian}, {Percival}, {Rhodes}, {Saavedra Criado},
  {Sauvage}, {Scaramella}, {Valenziano}, {Warren}, {Bender}, {Castander},
  {Cimatti}, {Le F{\`e}vre}, {Kurki-Suonio}, {Levi}, {Lilje}, {Meylan},
  {Nichol}, {Pedersen}, {Popa}, {Rebolo Lopez}, {Rix}, {Rottgering},
  {Zeilinger}, {Grupp}, {Hudelot}, {Massey}, {Meneghetti}, {Miller}, {Paltani},
  {Paulin-Henriksson}, {Pires}, {Saxton}, {Schrabback}, {Seidel}, {Walsh},
  {Aghanim}, {Amendola}, {Bartlett}, {Baccigalupi}, {Beaulieu}, {Benabed},
  {Cuby}, {Elbaz}, {Fosalba}, {Gavazzi}, {Helmi}, {Hook}, {Irwin}, {Kneib},
  {Kunz}, {Mannucci}, {Moscardini}, {Tao}, {Teyssier}, {Weller}, {Zamorani},
  {Zapatero Osorio}, {Boulade}, {Foumond}, {Di Giorgio}, {Guttridge}, {James},
  {Kemp}, {Martignac}, {Spencer}, {Walton}, {Bl{\"u}mchen}, {Bonoli},
  {Bortoletto}, {Cerna}, {Corcione}, {Fabron}, {Jahnke}, {Ligori}, {Madrid},
  {Martin}, {Morgante}, {Pamplona}, {Prieto}, {Riva}, {Toledo}, {Trifoglio},
  {Zerbi}, {Abdalla}, {Douspis}, {Grenet}, {Borgani}, {Bouwens}, {Courbin},
  {Delouis}, {Dubath}, {Fontana}, {Frailis}, {Grazian}, {Koppenh{\"o}fer},
  {Mansutti}, {Melchior}, {Mignoli}, {Mohr}, {Neissner}, {Noddle}, {Poncet},
  {Scodeggio}, {Serrano}, {Shane}, {Starck}, {Surace}, {Taylor},
  {Verdoes-Kleijn}, {Vuerli}, {Williams}, {Zacchei}, {Altieri}, {Escudero
  Sanz}, {Kohley}, {Oosterbroek}, {Astier}, {Bacon}, {Bardelli}, {Baugh},
  {Bellagamba}, {Benoist}, {Bianchi}, {Biviano}, {Branchini}, {Carbone},
  {Cardone}, {Clements}, {Colombi}, {Conselice}, {Cresci}, {Deacon}, {Dunlop},
  {Fedeli}, {Fontanot}, {Franzetti}, {Giocoli}, {Garcia-Bellido}, {Gow},
  {Heavens}, {Hewett}, {Heymans}, {Holland}, {Huang}, {Ilbert}, {Joachimi},
  {Jennins}, {Kerins}, {Kiessling}, {Kirk}, {Kotak}, {Krause}, {Lahav}, {van
  Leeuwen}, {Lesgourgues}, {Lombardi}, {Magliocchetti}, {Maguire}, {Majerotto},
  {Maoli}, {Marulli}, {Maurogordato}, {McCracken}, {McLure}, {Melchiorri},
  {Merson}, {Moresco}, {Nonino}, {Norberg}, {Peacock}, {Pello}, {Penny},
  {Pettorino}, {Di Porto}, {Pozzetti}, {Quercellini}, {Radovich}, {Rassat},
  {Roche}, {Ronayette}, {Rossetti}, {Sartoris}, {Schneider}, {Semboloni},
  {Serjeant}, {Simpson}, {Skordis}, {Smadja}, {Smartt}, {Spano}, {Spiro},
  {Sullivan}, {Tilquin}, {Trotta}, {Verde}, {Wang}, {Williger}, {Zhao},
  {Zoubian}, \& {Zucca}}]{laureijs2011}
{Laureijs}, R., {Amiaux}, J., {Arduini}, S., {et~al.} 2011, arXiv e-prints,
  arXiv:1110.3193.
\newblock \doarXiv{1110.3193}

\bibitem[{{Lazeyras} {et~al.}(2016){Lazeyras}, {Wagner}, {Baldauf}, \&
  {Schmidt}}]{lazeyras2016}
{Lazeyras}, T., {Wagner}, C., {Baldauf}, T., \& {Schmidt}, F. 2016, \jcap,
  2016, 018, \dodoi{10.1088/1475-7516/2016/02/018}

\bibitem[{{Leauthaud} {et~al.}(2016){Leauthaud}, {Bundy}, {Saito}, {Tinker},
  {Maraston}, {Tojeiro}, {Huang}, {Brownstein}, {Schneider}, \&
  {Thomas}}]{leauthaud2016}
{Leauthaud}, A., {Bundy}, K., {Saito}, S., {et~al.} 2016, \mnras, 457, 4021,
  \dodoi{10.1093/mnras/stw117}

\bibitem[{{Leistedt} \& {Peiris}(2014)}]{leistedt2014}
{Leistedt}, B., \& {Peiris}, H.~V. 2014, \mnras, 444, 2,
  \dodoi{10.1093/mnras/stu1439}

\bibitem[{{Lesci} {et~al.}(2020){Lesci}, {Marulli}, {Moscardini}, {Sereno},
  {Veropalumbo}, {Maturi}, {Giocoli}, {Radovich}, {Bellagamba}, {Roncarelli},
  {Bardelli}, {Contarini}, {Covone}, {Ingoglia}, {Nanni}, \&
  {Puddu}}]{lesci2020}
{Lesci}, G.~F., {Marulli}, F., {Moscardini}, L., {et~al.} 2020, arXiv e-prints,
  arXiv:2012.12273.
\newblock \doarXiv{2012.12273}

\bibitem[{{Lewis} {et~al.}(2000){Lewis}, {Challinor}, \& {Lasenby}}]{lewis2000}
{Lewis}, A., {Challinor}, A., \& {Lasenby}, A. 2000, \apj, 538, 473,
  \dodoi{10.1086/309179}

\bibitem[{{Linder}(2017)}]{linder2017}
{Linder}, E.~V. 2017, Astroparticle Physics, 86, 41,
  \dodoi{10.1016/j.astropartphys.2016.11.002}

\bibitem[{{LSST Dark Energy Science Collaboration}(2012)}]{LSST2012}
{LSST Dark Energy Science Collaboration}. 2012, arXiv e-prints,
  arXiv:1211.0310.
\newblock \doarXiv{1211.0310}

\bibitem[{{Maartens} \& {Majerotto}(2006)}]{maartens2006}
{Maartens}, R., \& {Majerotto}, E. 2006, \prd, 74, 023004,
  \dodoi{10.1103/PhysRevD.74.023004}

\bibitem[{{Majumdar} \& {Mohr}(2004)}]{majumdar2004}
{Majumdar}, S., \& {Mohr}, J.~J. 2004, \apj, 613, 41, \dodoi{10.1086/422829}

\bibitem[{{Mana} {et~al.}(2013){Mana}, {Giannantonio}, {Weller}, {Hoyle},
  {H{\"u}tsi}, \& {Sartoris}}]{mana2013}
{Mana}, A., {Giannantonio}, T., {Weller}, J., {et~al.} 2013, \mnras, 434, 684,
  \dodoi{10.1093/mnras/stt1062}

\bibitem[{{Marulli} {et~al.}(2012){Marulli}, {Bianchi}, {Branchini}, {Guzzo},
  {Moscardini}, \& {Angulo}}]{marulli2012b}
{Marulli}, F., {Bianchi}, D., {Branchini}, E., {et~al.} 2012, \mnras, 426,
  2566, \dodoi{10.1111/j.1365-2966.2012.21875.x}

\bibitem[{{Marulli} {et~al.}(2016){Marulli}, {Veropalumbo}, \&
  {Moresco}}]{marulli2016}
{Marulli}, F., {Veropalumbo}, A., \& {Moresco}, M. 2016, Astronomy and
  Computing, 14, 35, \dodoi{10.1016/j.ascom.2016.01.005}

\bibitem[{{Marulli} {et~al.}(2017){Marulli}, {Veropalumbo}, {Moscardini},
  {Cimatti}, \& {Dolag}}]{marulli2017}
{Marulli}, F., {Veropalumbo}, A., {Moscardini}, L., {Cimatti}, A., \& {Dolag},
  K. 2017, \aap, 599, A106, \dodoi{10.1051/0004-6361/201526885}

\bibitem[{{Marulli} {et~al.}(2018){Marulli}, {Veropalumbo}, {Sereno},
  {Moscardini}, {Pacaud}, {Pierre}, {Plionis}, {Cappi}, {Adami}, {Alis},
  {Altieri}, {Birkinshaw}, {Ettori}, {Faccioli}, {Gastaldello}, {Koulouridis},
  {Lidman}, {Le F{\`e}vre}, {Maurogordato}, {Poggianti}, {Pompei},
  {Sadibekova}, \& {Valtchanov}}]{marulli2018}
{Marulli}, F., {Veropalumbo}, A., {Sereno}, M., {et~al.} 2018, \aap, 620, A1,
  \dodoi{10.1051/0004-6361/201833238}

\bibitem[{{McDonald} \& {Roy}(2009)}]{mcdonald2009b}
{McDonald}, P., \& {Roy}, A. 2009, \jcap, 2009, 020,
  \dodoi{10.1088/1475-7516/2009/08/020}

\bibitem[{{Merloni} {et~al.}(2012){Merloni}, {Predehl}, {Becker},
  {B{\"o}hringer}, {Boller}, {Brunner}, {Brusa}, {Dennerl}, {Freyberg},
  {Friedrich}, {Georgakakis}, {Haberl}, {Hasinger}, {Meidinger}, {Mohr},
  {Nandra}, {Rau}, {Reiprich}, {Robrade}, {Salvato}, {Santangelo}, {Sasaki},
  {Schwope}, {Wilms}, \& {German eROSITA Consortium}}]{merloni2012}
{Merloni}, A., {Predehl}, P., {Becker}, W., {et~al.} 2012, arXiv e-prints,
  arXiv:1209.3114.
\newblock \doarXiv{1209.3114}

\bibitem[{{Miller} \& {Batuski}(2001)}]{miller2001}
{Miller}, C.~J., \& {Batuski}, D.~J. 2001, \apj, 551, 635,
  \dodoi{10.1086/320213}

\bibitem[{{Mohammad} {et~al.}(2018){Mohammad}, {Granett}, {Guzzo}, {Bel},
  {Branchini}, {de la Torre}, {Moscardini}, {Peacock}, {Bolzonella}, {Garilli},
  {Scodeggio}, {Abbas}, {Adami}, {Bottini}, {Cappi}, {Cucciati}, {Davidzon},
  {Franzetti}, {Fritz}, {Iovino}, {Krywult}, {Le Brun}, {Le F{\`e}vre},
  {Maccagni}, {Ma{\l}ek}, {Marulli}, {Polletta}, {Pollo}, {Tasca}, {Tojeiro},
  {Vergani}, {Zanichelli}, {Arnouts}, {Coupon}, {De Lucia}, {Ilbert}, \&
  {Moutard}}]{mohammad2018}
{Mohammad}, F.~G., {Granett}, B.~R., {Guzzo}, L., {et~al.} 2018, \aap, 610,
  A59, \dodoi{10.1051/0004-6361/201731685}

\bibitem[{{Moresco} \& {Marulli}(2017)}]{moresco2017}
{Moresco}, M., \& {Marulli}, F. 2017, \mnras, 471, L82,
  \dodoi{10.1093/mnrasl/slx112}

\bibitem[{{Moresco} {et~al.}(2021){Moresco}, {Veropalumbo}, {Marulli},
  {Moscardini}, \& {Cimatti}}]{moresco2021}
{Moresco}, M., {Veropalumbo}, A., {Marulli}, F., {Moscardini}, L., \&
  {Cimatti}, A. 2021, \apj, 919, 144, \dodoi{10.3847/1538-4357/ac10c9}

\bibitem[{{Moscardini} {et~al.}(2000){Moscardini}, {Matarrese}, {De Grandi}, \&
  {Lucchin}}]{moscardini2000b}
{Moscardini}, L., {Matarrese}, S., {De Grandi}, S., \& {Lucchin}, F. 2000,
  \mnras, 314, 647, \dodoi{10.1046/j.1365-8711.2000.03372.x}

\bibitem[{{Nadathur} {et~al.}(2019){Nadathur}, {Carter}, {Percival}, {Winther},
  \& {Bautista}}]{nadathur2019}
{Nadathur}, S., {Carter}, P.~M., {Percival}, W.~J., {Winther}, H.~A., \&
  {Bautista}, J.~E. 2019, \prd, 100, 023504,
  \dodoi{10.1103/PhysRevD.100.023504}

\bibitem[{{Nadathur} {et~al.}(2020){Nadathur}, {Woodfinden}, {Percival},
  {Aubert}, {Bautista}, {Dawson}, {Escoffier}, {Fromenteau}, {Gil-Mar{\'\i}n},
  {Rich}, {Ross}, {Rossi}, {Maga{\~n}a}, {Brownstein}, \&
  {Schneider}}]{nadathur2020}
{Nadathur}, S., {Woodfinden}, A., {Percival}, W.~J., {et~al.} 2020, \mnras,
  499, 4140, \dodoi{10.1093/mnras/staa3074}

\bibitem[{{Nanni} {et~al.}(in prep.)}]{nanni2020}
{Nanni}, L., {et~al.} in prep.

\bibitem[{{Neveux} {et~al.}(2020){Neveux}, {Burtin}, {de Mattia}, {Smith},
  {Ross}, {Hou}, {Bautista}, {Brinkmann}, {Chuang}, {Dawson}, {Gil-Mar{\'\i}n},
  {Lyke}, {de la Macorra}, {du Mas des Bourboux}, {Mohammad}, {M{\"u}ller},
  {Myers}, {Newman}, {Percival}, {Rossi}, {Schneider}, {Vivek}, {Zarrouk},
  {Zhao}, \& {Zhao}}]{neveux2020}
{Neveux}, R., {Burtin}, E., {de Mattia}, A., {et~al.} 2020, \mnras, 499, 210,
  \dodoi{10.1093/mnras/staa2780}

\bibitem[{{Okumura} {et~al.}(2016){Okumura}, {Hikage}, {Totani}, {Tonegawa},
  {Okada}, {Glazebrook}, {Blake}, {Ferreira}, {More}, \&
  {Taruya}}]{okumura2016}
{Okumura}, T., {Hikage}, C., {Totani}, T., {et~al.} 2016, \pasj, 68, 38,
  \dodoi{10.1093/pasj/psw029}

\bibitem[{{Pacaud} {et~al.}(2018){Pacaud}, {Pierre}, {Melin}, {Adami},
  {Evrard}, {Galli}, {Gastaldello}, {Maughan}, {Sereno}, {Alis}, {Altieri},
  {Birkinshaw}, {Chiappetti}, {Faccioli}, {Giles}, {Horellou}, {Iovino},
  {Koulouridis}, {Le F{\`e}vre}, {Lidman}, {Lieu}, {Maurogordato},
  {Moscardini}, {Plionis}, {Poggianti}, {Pompei}, {Sadibekova}, {Valtchanov},
  \& {Willis}}]{pacaud2018}
{Pacaud}, F., {Pierre}, M., {Melin}, J.~B., {et~al.} 2018, \aap, 620, A10,
  \dodoi{10.1051/0004-6361/201834022}

\bibitem[{{Peacock} \& {Dodds}(1996)}]{peacock1996}
{Peacock}, J.~A., \& {Dodds}, S.~J. 1996, \mnras, 280, L19

\bibitem[{{Peacock} {et~al.}(2001){Peacock}, {Cole}, {Norberg}, {Baugh},
  {Bland-Hawthorn}, {Bridges}, {Cannon}, {Colless}, {Collins}, {Couch},
  {Dalton}, {Deeley}, {De Propris}, {Driver}, {Efstathiou}, {Ellis}, {Frenk},
  {Glazebrook}, {Jackson}, {Lahav}, {Lewis}, {Lumsden}, {Maddox}, {Percival},
  {Peterson}, {Price}, {Sutherland}, \& {Taylor}}]{peacock2001}
{Peacock}, J.~A., {Cole}, S., {Norberg}, P., {et~al.} 2001, \nat, 410, 169.
\newblock \doarXiv{astro-ph/0103143}

\bibitem[{{Percival} {et~al.}(2004){Percival}, {Burkey}, {Heavens}, {Taylor},
  {Cole}, {Peacock}, {Baugh}, {Bland-Hawthorn}, {Bridges}, {Cannon}, {Colless},
  {Collins}, {Couch}, {Dalton}, {De Propris}, {Driver}, {Efstathiou}, {Ellis},
  {Frenk}, {Glazebrook}, {Jackson}, {Lahav}, {Lewis}, {Lumsden}, {Maddox},
  {Norberg}, {Peterson}, {Sutherland}, \& {Taylor}}]{percival2004}
{Percival}, W.~J., {Burkey}, D., {Heavens}, A., {et~al.} 2004, \mnras, 353,
  1201, \dodoi{10.1111/j.1365-2966.2004.08146.x}

\bibitem[{{Percival} {et~al.}(2014){Percival}, {Ross}, {S{\'a}nchez},
  {Samushia}, {Burden}, {Crittenden}, {Cuesta}, {Magana}, {Manera}, {Beutler},
  {Chuang}, {Eisenstein}, {Ho}, {McBride}, {Montesano}, {Padmanabhan}, {Reid},
  {Saito}, {Schneider}, {Seo}, {Tojeiro}, \& {Weaver}}]{percival2014}
{Percival}, W.~J., {Ross}, A.~J., {S{\'a}nchez}, A.~G., {et~al.} 2014, \mnras,
  439, 2531, \dodoi{10.1093/mnras/stu112}

\bibitem[{{Pezzotta} {et~al.}(2017){Pezzotta}, {de la Torre}, {Bel}, {Granett},
  {Guzzo}, {Peacock}, {Garilli}, {Scodeggio}, {Bolzonella}, {Abbas}, {Adami},
  {Bottini}, {Cappi}, {Cucciati}, {Davidzon}, {Franzetti}, {Fritz}, {Iovino},
  {Krywult}, {Le Brun}, {Le F{\`e}vre}, {Maccagni}, {Ma{\l}ek}, {Marulli},
  {Polletta}, {Pollo}, {Tasca}, {Tojeiro}, {Vergani}, {Zanichelli}, {Arnouts},
  {Branchini}, {Coupon}, {De Lucia}, {Koda}, {Ilbert}, {Mohammad}, {Moutard},
  \& {Moscardini}}]{pezzotta2017}
{Pezzotta}, A., {de la Torre}, S., {Bel}, J., {et~al.} 2017, \aap, 604, A33,
  \dodoi{10.1051/0004-6361/201630295}

\bibitem[{{Planck Collaboration} {et~al.}(2020){Planck Collaboration},
  {Aghanim}, {Akrami}, {Ashdown}, {Aumont}, {Baccigalupi}, {Ballardini},
  {Banday}, {Barreiro}, {Bartolo}, {Basak}, {Battye}, {Benabed}, {Bernard},
  {Bersanelli}, {Bielewicz}, {Bock}, {Bond}, {Borrill}, {Bouchet}, {Boulanger},
  {Bucher}, {Burigana}, {Butler}, {Calabrese}, {Cardoso}, {Carron},
  {Challinor}, {Chiang}, {Chluba}, {Colombo}, {Combet}, {Contreras}, {Crill},
  {Cuttaia}, {de Bernardis}, {de Zotti}, {Delabrouille}, {Delouis}, {Di
  Valentino}, {Diego}, {Dor{\'e}}, {Douspis}, {Ducout}, {Dupac}, {Dusini},
  {Efstathiou}, {Elsner}, {En{\ss}lin}, {Eriksen}, {Fantaye}, {Farhang},
  {Fergusson}, {Fernandez-Cobos}, {Finelli}, {Forastieri}, {Frailis},
  {Fraisse}, {Franceschi}, {Frolov}, {Galeotta}, {Galli}, {Ganga},
  {G{\'e}nova-Santos}, {Gerbino}, {Ghosh}, {Gonz{\'a}lez-Nuevo}, {G{\'o}rski},
  {Gratton}, {Gruppuso}, {Gudmundsson}, {Hamann}, {Handley}, {Hansen},
  {Herranz}, {Hildebrandt}, {Hivon}, {Huang}, {Jaffe}, {Jones}, {Karakci},
  {Keih{\"a}nen}, {Keskitalo}, {Kiiveri}, {Kim}, {Kisner}, {Knox},
  {Krachmalnicoff}, {Kunz}, {Kurki-Suonio}, {Lagache}, {Lamarre}, {Lasenby},
  {Lattanzi}, {Lawrence}, {Le Jeune}, {Lemos}, {Lesgourgues}, {Levrier},
  {Lewis}, {Liguori}, {Lilje}, {Lilley}, {Lindholm}, {L{\'o}pez-Caniego},
  {Lubin}, {Ma}, {Mac{\'\i}as-P{\'e}rez}, {Maggio}, {Maino}, {Mandolesi},
  {Mangilli}, {Marcos-Caballero}, {Maris}, {Martin}, {Martinelli},
  {Mart{\'\i}nez-Gonz{\'a}lez}, {Matarrese}, {Mauri}, {McEwen}, {Meinhold},
  {Melchiorri}, {Mennella}, {Migliaccio}, {Millea}, {Mitra},
  {Miville-Desch{\^e}nes}, {Molinari}, {Montier}, {Morgante}, {Moss}, {Natoli},
  {N{\o}rgaard-Nielsen}, {Pagano}, {Paoletti}, {Partridge}, {Patanchon},
  {Peiris}, {Perrotta}, {Pettorino}, {Piacentini}, {Polastri}, {Polenta},
  {Puget}, {Rachen}, {Reinecke}, {Remazeilles}, {Renzi}, {Rocha}, {Rosset},
  {Roudier}, {Rubi{\~n}o-Mart{\'\i}n}, {Ruiz-Granados}, {Salvati}, {Sandri},
  {Savelainen}, {Scott}, {Shellard}, {Sirignano}, {Sirri}, {Spencer},
  {Sunyaev}, {Suur-Uski}, {Tauber}, {Tavagnacco}, {Tenti}, {Toffolatti},
  {Tomasi}, {Trombetti}, {Valenziano}, {Valiviita}, {Van Tent}, {Vibert},
  {Vielva}, {Villa}, {Vittorio}, {Wandelt}, {Wehus}, {White}, {White},
  {Zacchei}, \& {Zonca}}]{Planck2018}
{Planck Collaboration}, {Aghanim}, N., {Akrami}, Y., {et~al.} 2020, \aap, 641,
  A6, \dodoi{10.1051/0004-6361/201833910}

\bibitem[{{Reid} {et~al.}(2016){Reid}, {Ho}, {Padmanabhan}, {Percival},
  {Tinker}, {Tojeiro}, {White}, {Eisenstein}, {Maraston}, {Ross},
  {S{\'a}nchez}, {Schlegel}, {Sheldon}, {Strauss}, {Thomas}, {Wake}, {Beutler},
  {Bizyaev}, {Bolton}, {Brownstein}, {Chuang}, {Dawson}, {Harding}, {Kitaura},
  {Leauthaud}, {Masters}, {McBride}, {More}, {Olmstead}, {Oravetz}, {Nuza},
  {Pan}, {Parejko}, {Pforr}, {Prada}, {Rodr{\'\i}guez-Torres},
  {Salazar-Albornoz}, {Samushia}, {Schneider}, {Sc{\'o}ccola}, {Simmons}, \&
  {Vargas-Magana}}]{reid2016}
{Reid}, B., {Ho}, S., {Padmanabhan}, N., {et~al.} 2016, \mnras, 455, 1553,
  \dodoi{10.1093/mnras/stv2382}

\bibitem[{{Reid} {et~al.}(2012){Reid}, {Samushia}, {White}, {Percival},
  {Manera}, {Padmanabhan}, {Ross}, {S{\'a}nchez}, {Bailey}, {Bizyaev},
  {Bolton}, {Brewington}, {Brinkmann}, {Brownstein}, {Cuesta}, {Eisenstein},
  {Gunn}, {Honscheid}, {Malanushenko}, {Malanushenko}, {Maraston}, {McBride},
  {Muna}, {Nichol}, {Oravetz}, {Pan}, {de Putter}, {Roe}, {Ross}, {Schlegel},
  {Schneider}, {Seo}, {Shelden}, {Sheldon}, {Simmons}, {Skibba}, {Snedden},
  {Swanson}, {Thomas}, {Tinker}, {Tojeiro}, {Verde}, {Wake}, {Weaver},
  {Weinberg}, {Zehavi}, \& {Zhao}}]{reid2012}
{Reid}, B.~A., {Samushia}, L., {White}, M., {et~al.} 2012, \mnras, 426, 2719,
  \dodoi{10.1111/j.1365-2966.2012.21779.x}

\bibitem[{{Ross} {et~al.}(2017){Ross}, {Beutler}, {Chuang}, {Pellejero-Ibanez},
  {Seo}, {Vargas-Maga{\~n}a}, {Cuesta}, {Percival}, {Burden}, {S{\'a}nchez},
  {Grieb}, {Reid}, {Brownstein}, {Dawson}, {Eisenstein}, {Ho}, {Kitaura},
  {Nichol}, {Olmstead}, {Prada}, {Rodr{\'\i}guez-Torres}, {Saito},
  {Salazar-Albornoz}, {Schneider}, {Thomas}, {Tinker}, {Tojeiro}, {Wang},
  {White}, \& {Zhao}}]{ross2017}
{Ross}, A.~J., {Beutler}, F., {Chuang}, C.-H., {et~al.} 2017, \mnras, 464,
  1168, \dodoi{10.1093/mnras/stw2372}

\bibitem[{{Saito} {et~al.}(2014){Saito}, {Baldauf}, {Vlah}, {Seljak},
  {Okumura}, \& {McDonald}}]{saito2014}
{Saito}, S., {Baldauf}, T., {Vlah}, Z., {et~al.} 2014, \prd, 90, 123522,
  \dodoi{10.1103/PhysRevD.90.123522}

\bibitem[{{Saito} {et~al.}(2016){Saito}, {Leauthaud}, {Hearin}, {Bundy},
  {Zentner}, {Behroozi}, {Reid}, {Sinha}, {Coupon}, {Tinker}, {White}, \&
  {Schneider}}]{saito2016}
{Saito}, S., {Leauthaud}, A., {Hearin}, A.~P., {et~al.} 2016, \mnras, 460,
  1457, \dodoi{10.1093/mnras/stw1080}

\bibitem[{{Samushia} {et~al.}(2012){Samushia}, {Percival}, \&
  {Raccanelli}}]{samushia2012}
{Samushia}, L., {Percival}, W.~J., \& {Raccanelli}, A. 2012, \mnras, 420, 2102,
  \dodoi{10.1111/j.1365-2966.2011.20169.x}

\bibitem[{{Samushia} {et~al.}(2014){Samushia}, {Reid}, {White}, {Percival},
  {Cuesta}, {Zhao}, {Ross}, {Manera}, {Aubourg}, \& {Beutler}}]{samushia2014}
{Samushia}, L., {Reid}, B.~A., {White}, M., {et~al.} 2014, \mnras, 439, 3504,
  \dodoi{10.1093/mnras/stu197}

\bibitem[{{S{\'a}nchez}(2020)}]{sanchez2020}
{S{\'a}nchez}, A.~G. 2020, \prd, 102, 123511,
  \dodoi{10.1103/PhysRevD.102.123511}

\bibitem[{{Sartoris} {et~al.}(2016){Sartoris}, {Biviano}, {Fedeli}, {Bartlett},
  {Borgani}, {Costanzi}, {Giocoli}, {Moscardini}, {Weller}, {Ascaso},
  {Bardelli}, {Maurogordato}, \& {Viana}}]{sartoris2016}
{Sartoris}, B., {Biviano}, A., {Fedeli}, C., {et~al.} 2016, \mnras, 459, 1764,
  \dodoi{10.1093/mnras/stw630}

\bibitem[{{Schuecker} {et~al.}(2003){Schuecker}, {B{\"o}hringer}, {Collins}, \&
  {Guzzo}}]{schuecker2003}
{Schuecker}, P., {B{\"o}hringer}, H., {Collins}, C.~A., \& {Guzzo}, L. 2003,
  \aap, 398, 867, \dodoi{10.1051/0004-6361:20021715}

\bibitem[{{Schuecker} {et~al.}(2001){Schuecker}, {B{\"o}hringer}, {Guzzo},
  {Collins}, {Neumann}, {Schindler}, {Voges}, {De Grandi}, {Chincarini},
  {Cruddace}, {M{\"u}ller}, {Reiprich}, {Retzlaff}, \&
  {Shaver}}]{schuecker2001}
{Schuecker}, P., {B{\"o}hringer}, H., {Guzzo}, L., {et~al.} 2001, \aap, 368,
  86, \dodoi{10.1051/0004-6361:20000542}

\bibitem[{{Scoccimarro}(2004)}]{scoccimarro2004}
{Scoccimarro}, R. 2004, \prd, 70, 083007, \dodoi{10.1103/PhysRevD.70.083007}

\bibitem[{{Sereno} {et~al.}(2015){Sereno}, {Veropalumbo}, {Marulli}, {Covone},
  {Moscardini}, \& {Cimatti}}]{sereno2015}
{Sereno}, M., {Veropalumbo}, A., {Marulli}, F., {et~al.} 2015, \mnras, 449,
  4147, \dodoi{10.1093/mnras/stv280}

\bibitem[{{Sheth} {et~al.}(2001){Sheth}, {Mo}, \&
  {Tormen}}]{sheth_mo_tormen2001}
{Sheth}, R.~K., {Mo}, H.~J., \& {Tormen}, G. 2001, \mnras, 323, 1,
  \dodoi{10.1046/j.1365-8711.2001.04006.x}

\bibitem[{{Spergel} {et~al.}(2015){Spergel}, {Gehrels}, {Baltay}, {Bennett},
  {Breckinridge}, {Donahue}, {Dressler}, {Gaudi}, {Greene}, {Guyon}, {Hirata},
  {Kalirai}, {Kasdin}, {Macintosh}, {Moos}, {Perlmutter}, {Postman},
  {Rauscher}, {Rhodes}, {Wang}, {Weinberg}, {Benford}, {Hudson}, {Jeong},
  {Mellier}, {Traub}, {Yamada}, {Capak}, {Colbert}, {Masters}, {Penny},
  {Savransky}, {Stern}, {Zimmerman}, {Barry}, {Bartusek}, {Carpenter}, {Cheng},
  {Content}, {Dekens}, {Demers}, {Grady}, {Jackson}, {Kuan}, {Kruk}, {Melton},
  {Nemati}, {Parvin}, {Poberezhskiy}, {Peddie}, {Ruffa}, {Wallace}, {Whipple},
  {Wollack}, \& {Zhao}}]{spergel2015}
{Spergel}, D., {Gehrels}, N., {Baltay}, C., {et~al.} 2015, arXiv e-prints,
  arXiv:1503.03757.
\newblock \doarXiv{1503.03757}

\bibitem[{{Sridhar} {et~al.}(2017){Sridhar}, {Maurogordato}, {Benoist},
  {Cappi}, \& {Marulli}}]{sridhar2017}
{Sridhar}, S., {Maurogordato}, S., {Benoist}, C., {Cappi}, A., \& {Marulli}, F.
  2017, \aap, 600, A32, \dodoi{10.1051/0004-6361/201629369}

\bibitem[{{Sugiyama} {et~al.}(2020){Sugiyama}, {Saito}, {Beutler}, \&
  {Seo}}]{sugiyama2020}
{Sugiyama}, N.~S., {Saito}, S., {Beutler}, F., \& {Seo}, H.-J. 2020, \mnras,
  497, 1684, \dodoi{10.1093/mnras/staa1940}

\bibitem[{{Swanson} {et~al.}(2008){Swanson}, {Tegmark}, {Hamilton}, \&
  {Hill}}]{swanson2008_MANGLE}
{Swanson}, M.~E.~C., {Tegmark}, M., {Hamilton}, A. J.~S., \& {Hill}, J.~C.
  2008, \mnras, 387, 1391, \dodoi{10.1111/j.1365-2966.2008.13296.x}

\bibitem[{{Tamone} {et~al.}(2020){Tamone}, {Raichoor}, {Zhao}, {de Mattia},
  {Gorgoni}, {Burtin}, {Ruhlmann-Kleider}, {Ross}, {Alam}, {Percival}, {Avila},
  {Chapman}, {Chuang}, {Comparat}, {Dawson}, {de la Torre}, {du Mas des
  Bourboux}, {Escoffier}, {Gonzalez-Perez}, {Hou}, {Kneib}, {Mohammad},
  {Mueller}, {Paviot}, {Rossi}, {Schneider}, {Wang}, \& {Zhao}}]{tamone2020}
{Tamone}, A., {Raichoor}, A., {Zhao}, C., {et~al.} 2020, \mnras, 499, 5527,
  \dodoi{10.1093/mnras/staa3050}

\bibitem[{{Taruya} \& {Hiramatsu}(2008)}]{taruya2008}
{Taruya}, A., \& {Hiramatsu}, T. 2008, \apj, 674, 617, \dodoi{10.1086/526515}

\bibitem[{{Taruya} {et~al.}(2010){Taruya}, {Nishimichi}, \&
  {Saito}}]{taruya2010}
{Taruya}, A., {Nishimichi}, T., \& {Saito}, S. 2010, \prd, 82, 063522,
  \dodoi{10.1103/PhysRevD.82.063522}

\bibitem[{{Taruya} {et~al.}(2011){Taruya}, {Saito}, \&
  {Nishimichi}}]{taruya2011}
{Taruya}, A., {Saito}, S., \& {Nishimichi}, T. 2011, \prd, 83, 103527,
  \dodoi{10.1103/PhysRevD.83.103527}

\bibitem[{{Tempel} {et~al.}(2014){Tempel}, {Tamm}, {Gramann}, {Tuvikene},
  {Liivam{\"a}gi}, {Suhhonenko}, {Kipper}, {Einasto}, \& {Saar}}]{tempel2014}
{Tempel}, E., {Tamm}, A., {Gramann}, M., {et~al.} 2014, \aap, 566, A1,
  \dodoi{10.1051/0004-6361/201423585}

\bibitem[{{Tinker} {et~al.}(2008){Tinker}, {Kravtsov}, {Klypin}, {Abazajian},
  {Warren}, {Yepes}, {Gottl{\"o}ber}, \& {Holz}}]{tinker2008}
{Tinker}, J., {Kravtsov}, A.~V., {Klypin}, A., {et~al.} 2008, \apj, 688, 709,
  \dodoi{10.1086/591439}

\bibitem[{{Tinker} {et~al.}(2010){Tinker}, {Robertson}, {Kravtsov}, {Klypin},
  {Warren}, {Yepes}, \& {Gottl{\"o}ber}}]{tinker2010}
{Tinker}, J.~L., {Robertson}, B.~E., {Kravtsov}, A.~V., {et~al.} 2010, \apj,
  724, 878, \dodoi{10.1088/0004-637X/724/2/878}

\bibitem[{{Tojeiro} {et~al.}(2012){Tojeiro}, {Percival}, {Brinkmann},
  {Brownstein}, {Eisenstein}, {Manera}, {Maraston}, {McBride}, {Muna}, {Reid},
  {Ross}, {Ross}, {Samushia}, {Padmanabhan}, {Schneider}, {Skibba},
  {S{\'a}nchez}, {Swanson}, {Thomas}, {Tinker}, {Verde}, {Wake}, {Weaver}, \&
  {Zhao}}]{tojeiro2012}
{Tojeiro}, R., {Percival}, W.~J., {Brinkmann}, J., {et~al.} 2012, \mnras, 424,
  2339, \dodoi{10.1111/j.1365-2966.2012.21404.x}

\bibitem[{{Turnbull} {et~al.}(2012){Turnbull}, {Hudson}, {Feldman}, {Hicken},
  {Kirshner}, \& {Watkins}}]{turnbull2012}
{Turnbull}, S.~J., {Hudson}, M.~J., {Feldman}, H.~A., {et~al.} 2012, \mnras,
  420, 447, \dodoi{10.1111/j.1365-2966.2011.20050.x}

\bibitem[{{Valageas} \& {Clerc}(2012)}]{valageas2012}
{Valageas}, P., \& {Clerc}, N. 2012, \aap, 547, A100,
  \dodoi{10.1051/0004-6361/201219646}

\bibitem[{{Veropalumbo} {et~al.}(2014){Veropalumbo}, {Marulli}, {Moscardini},
  {Moresco}, \& {Cimatti}}]{veropalumbo2014}
{Veropalumbo}, A., {Marulli}, F., {Moscardini}, L., {Moresco}, M., \&
  {Cimatti}, A. 2014, \mnras, 442, 3275, \dodoi{10.1093/mnras/stu1050}

\bibitem[{{Veropalumbo} {et~al.}(2016){Veropalumbo}, {Marulli}, {Moscardini},
  {Moresco}, \& {Cimatti}}]{veropalumbo2016}
---. 2016, \mnras, 458, 1909, \dodoi{10.1093/mnras/stw306}

\bibitem[{{Veropalumbo} {et~al.}(in prep.)}]{veropalumbo2021}
{Veropalumbo}, A., {et~al.} in prep.

\bibitem[{{Vikhlinin} {et~al.}(2009){Vikhlinin}, {Kravtsov}, {Burenin},
  {Ebeling}, {Forman}, {Hornstrup}, {Jones}, {Murray}, {Nagai}, {Quintana}, \&
  {Voevodkin}}]{vikhlinin2009}
{Vikhlinin}, A., {Kravtsov}, A.~V., {Burenin}, R.~A., {et~al.} 2009, \apj, 692,
  1060, \dodoi{10.1088/0004-637X/692/2/1060}

\bibitem[{{Wang} {et~al.}(2020){Wang}, {Zhao}, {Zhao}, {Philcox}, {Alam},
  {Tamone}, {de Mattia}, {Ross}, {Raichoor}, {Burtin}, {Paviot}, {de la Torre},
  {Percival}, {Dawson}, {Gil-Mar{\'\i}n}, {Bautista}, {Hou}, {Koyama},
  {Peacock}, {Ruhlmann-Kleider}, {Bourboux}, {Chuang}, {Comparat}, {Escoffier},
  {Kneib}, {Mueller}, {Newman}, {Rossi}, {Shafieloo}, \&
  {Schneider}}]{wang2020}
{Wang}, Y., {Zhao}, G.-B., {Zhao}, C., {et~al.} 2020, \mnras, 498, 3470,
  \dodoi{10.1093/mnras/staa2593}

\bibitem[{{Wen} {et~al.}(2010){Wen}, {Han}, \& {Liu}}]{wen2010}
{Wen}, Z.~L., {Han}, J.~L., \& {Liu}, F.~S. 2010, \mnras, 407, 533,
  \dodoi{10.1111/j.1365-2966.2010.16930.x}

\bibitem[{{Wen} {et~al.}(2012){Wen}, {Han}, \& {Liu}}]{wen2012}
---. 2012, \apjs, 199, 34, \dodoi{10.1088/0067-0049/199/2/34}

\bibitem[{{Wetterich}(1995)}]{wetterich1995}
{Wetterich}, C. 1995, \aap, 301, 321

\bibitem[{{Zehavi} {et~al.}(2011){Zehavi}, {Zheng}, {Weinberg}, {Blanton},
  {Bahcall}, {Berlind}, {Brinkmann}, {Frieman}, {Gunn}, {Lupton}, {Nichol},
  {Percival}, {Schneider}, {Skibba}, {Strauss}, {Tegmark}, \&
  {York}}]{zehavi2011}
{Zehavi}, I., {Zheng}, Z., {Weinberg}, D.~H., {et~al.} 2011, \apj, 736, 59,
  \dodoi{10.1088/0004-637X/736/1/59}

\bibitem[{{Zhang} {et~al.}(2008){Zhang}, {Yu}, {Noh}, \& {Zhu}}]{zhang2008}
{Zhang}, H., {Yu}, H., {Noh}, H., \& {Zhu}, Z.-H. 2008, Physics Letters B, 665,
  319, \dodoi{10.1016/j.physletb.2008.06.041}

\bibitem[{{Zhao} {et~al.}(2021){Zhao}, {Wang}, {Taruya}, {Zhang},
  {Gil-Mar{\'\i}n}, {de Mattia}, {Ross}, {Raichoor}, {Zhao}, {Percival},
  {Alam}, {Bautista}, {Burtin}, {Chuang}, {Dawson}, {Hou}, {Kneib}, {Koyama},
  {du Mas des Bourboux}, {Mueller}, {Newman}, {Peacock}, {Rossi},
  {Ruhlmann-Kleider}, {Schneider}, \& {Shafieloo}}]{zhao2021}
{Zhao}, G.-B., {Wang}, Y., {Taruya}, A., {et~al.} 2021, \mnras, 504, 33,
  \dodoi{10.1093/mnras/stab849}

\bibitem[{{Zurek} {et~al.}(1994){Zurek}, {Quinn}, {Salmon}, \&
  {Warren}}]{zurek1994}
{Zurek}, W.~H., {Quinn}, P.~J., {Salmon}, J.~K., \& {Warren}, M.~S. 1994, \apj,
  431, 559, \dodoi{10.1086/174507}

\end{thebibliography}
\bibliographystyle{aasjournal}

\end{document}